\documentclass[12pt]{article} 
\RequirePackage[colorlinks,citecolor=blue,urlcolor=blue]{hyperref}
\usepackage[utf8]{inputenc} 

\usepackage{geometry} 
\usepackage{kantlipsum}
\newlength{\normalparindent}
\raggedright
\usepackage{hyperref}

\usepackage{graphicx} 

\usepackage{booktabs} 
\usepackage{array} 
\usepackage{paralist} 
\usepackage{verbatim} 
\usepackage{subcaption} 
\usepackage{amsbsy}
\usepackage{amsmath}
\usepackage{amssymb}
\usepackage{amsfonts}
\usepackage{amscd}
\usepackage{amsthm}
\usepackage{float}
\usepackage[toc,page]{appendix}
\usepackage{natbib}
\usepackage{lineno}
\usepackage{mathrsfs}
\usepackage{bbm}
\usepackage{lineno}
\usepackage{todonotes}
\usepackage{setspace}
\usepackage{multirow}
\usepackage{todonotes}
\usepackage{fancyhdr} 
\usepackage{sectsty}
\allsectionsfont{\sffamily\mdseries\upshape} 

\usepackage[nottoc,notlof,notlot]{tocbibind} 
\usepackage[titles,subfigure]{tocloft} 

\newcommand{\onDryad}{in the Supplementary Material available on Dryad}

\newcommand{\prob}[1]{{\mathbb P}\hspace{-0.2em} \left(  #1 \right)}
\newcommand{\transpose}{'}

\newcommand{\data}{\mathbf{Y}}

\newcommand{\phylogeny}{{\cal F}}
\newcommand{\nTaxa}{N}

\newcommand{\nSites}{C}
\newcommand{\site}{c}

\newcommand{\category}{r}
\newcommand{\nCategories}{R}
\newcommand{\rateCategory}{\gamma}

\newcommand{\likelihoodAll}[1]{{\mathbb P}\hspace{-0.2em} \left( \data \mid #1 \right)}
\newcommand{\likelihood}[1]{{\mathbb P}\hspace{-0.2em} \left( \data_{\site} \mid #1 \right)}

\newcommand{\argmin}{\operatorname{arg\,min}}

\newcommand{\alphabetsize}{S}

\newcommand{\approxnabla}{\widetilde{\nabla}}

\newcommand{\param}{\theta}
\newcommand{\params}{\boldsymbol{\param}}
\newcommand{\randomeffect}{\epsilon}

\newcommand{\ratematrix}{\mathbf{Q}}
\newcommand{\ratematrixelement}{\lambda}
\newcommand{\branchlength}{t}

\newcommand{\mapping}{\mathbf{M}}
\newcommand{\allQderivs}{\mathbf{D}}
\newcommand{\allQderivselement}{d}

\newcommand{\node}{v}
\newcommand{\parent}{u}
\newcommand{\sister}{w}

\newcommand{\postPartial}{p}
\newcommand{\prePartial}{q}
\newcommand{\bpostPartial}{\mathbf{\postPartial}}
\newcommand{\bprePartial}{\mathbf{\prePartial}}
\newcommand{\bprePrePartial}{\mathbf{\tilde{\prePartial}}}

\newcommand{\vecm}{\operatorname{vec}}

\newcommand{\indicatormatrix}{\mathbf{E}}
\newcommand{\indicatormatrixij}{\mathbf{E}_{ij}}

\newcommand{\bA}{\mathbf{A}}
\newcommand{\bB}{\mathbf{B}}

\newcommand{\bp}{\mathbf{p}}
\newcommand{\bP}{\mathbf{P}}

\newcommand{\bQ}{\mathbf{Q}}

\newcommand{\bR}{\mathbf{R}}

\newcommand{\bS}{\mathbf{S}}

\newcommand{\bX}{\mathbf{X}}
\newcommand{\by}{\mathbf{y}}
\newcommand{\bY}{\mathbf{Y}}
\newcommand{\bZ}{\mathbf{Z}}

\newcommand{\basemodel}{\mathcal{M}}

\newcommand{\bigO}{\mathcal{O}}
\newcommand{\order}[1]{\bigO \hspace{-0.1em} \left( #1 \right)}
\newcommand{\mnorm}{\vert \vert}
\newcommand{\bigmnorm}{\Big\vert \Big\vert}
\newcommand{\Biggmnorm}{\Bigg\vert \Bigg\vert}

\newcommand{\globalScale}{\tau}
\newcommand{\bridgeExponent}{\alpha}

\newcommand{\bbeta}{\boldsymbol{\beta}}

\newcommand{\pr}{\text{Pr}}

\newcommand{\indicator}{\mathbb{I}}
\newcommand{\indicatorvector}{\boldsymbol{\indicator}}

\newcommand{\bpi}{\boldsymbol{\pi}}

\newcommand{\posteriorFunction}{f}

\newcommand{\perBranchParam}{\eta}

\newcommand{\dBydParam}{\frac{\partial}{\partial \param_k}}
\newcommand{\dBydPerBranchParam}{\frac{\partial}{\partial \perBranchParam_{\node k}}}

\newcommand{\rootFreqs}{\boldsymbol{\pi}_{\text{root}}}
\newcommand{\rootIndex}{2 \nTaxa - 1,}
\newcommand{\rootContribution}{\bpostPartial_{\rootIndex \category \site}' \dBydParam \rootFreqs}

\newcommand{\rootLabel}{R \hspace{-0.1em} \left( \theta_k \right)}

\newcommand{\mom}{\boldsymbol{\xi}}
\newcommand{\zero}{\boldsymbol{0}}
\newcommand{\cov}{\mathbf{M}}

\def\newshortcut#1#2{%
\let#1=\undefined
\newcommand{#1}{#2}}

\newshortcut{\pyabove}{\cDensity{\latentData_{\nodeIndexOne}}{\latentData_{\above{\nodeIndexOne}}}}

\renewcommand{\theequation}{\arabic{equation}}

\newcommand{\beginsupplement}{%
        \setcounter{table}{0}
        \renewcommand{\thetable}{S\arabic{table}}%
        \setcounter{figure}{0}
        \renewcommand{\thefigure}{S\arabic{figure}}%
        \renewcommand{\thesection}{S\arabic{section}}%
        \setcounter{section}{0}
        \setcounter{equation}{0}
        \renewcommand{\theequation}{S\arabic{equation}}
}

\begin{document}
\singlespacing
\begin{flushright}
Version dated: \today\\
\end{flushright}

\bigskip
\medskip
\begin{center}

\noindent{\Large \bf
	Random-effects substitution models for phylogenetics via scalable gradient approximations
}
\bigskip

\noindent{\normalsize \sc
	Andrew F.~Magee$^{1}$ \\
	Andrew J.~Holbrook$^{1}$ \\
	Jonathan E.~Pekar$^{2,3}$ \\
	Itzue W.~Caviedes-Solis$^{4}$ \\
	Fredrick A.~Matsen IV$^{5,6,7,8}$ \\
	Guy Baele$^{9}$ \\
	Joel O.~Wertheim$^{10}$ \\
	Xiang Ji$^{11}$ \\
	Philippe Lemey$^{9}$ \\
    and Marc A.~Suchard$^{1,12,13}$} \\

\bigskip
\noindent {\small
  \it $^1$ Department of Biostatistics, Jonathan and Karin Fielding School of Public Health, University of California Los Angeles, Los Angeles, CA, USA \\
  \it $^2$ Bioinformatics and Systems Biology Graduate Program, University of California San Diego, La Jolla, CA, USA \\
  \it $^3$ Department of Biomedical Informatics, University of California San Diega, La Jolla, CA, USA \\
  \it $^4$ Department of Biology, Swarthmore College, Swarthmore, PA, USA \\
  \it $^5$ Howard Hughes Medical Institute, Seattle, Washington, USA \\
  \it $^6$ Computational Biology Program, Fred Hutchinson Cancer Research Center, Seattle, Washington, USA \\
  \it $^7$ Department of Genome Sciences, University of Washington, Seattle, Washington, USA \\
  \it $^8$ Department of Statistics, University of Washington, Seattle, Washington, USA \\
  \it $^9$ Department of Microbiology, Immunology and Transplantation, Rega Institute, KU Leuven, Leuven, Belgium \\
  \it $^{10}$ Department of Medicine, University of California San Diego, La Jolla, CA, USA \\
  \it $^{11}$ Department of Mathematics, Tulane University, New Orleans, LA, USA \\
  \it $^{12}$Department of Biomathematics, David Geffen School of Medicine at UCLA, University of California Los Angeles, Los Angeles, CA, USA \\
  \it $^{13}$ Department of Human Genetics, David Geffen School of Medicine at UCLA, Universtiy of California Los Angeles, Los Angeles, CA, USA
} \\

\end{center}
\medskip
\noindent{\bf Corresponding author:} Marc A.~Suchard, Departments of Biostatistics, Biomathematics, and Human Genetics,
University of California Los Angeles, 695 Charles E.~Young Dr., South,
Los Angeles, CA 90095-7088, USA; E-mail: \url{msuchard@ucla.edu}

\vspace{1in}

\clearpage

\doublespacing

\paragraph{Abstract}
Phylogenetic and discrete-trait evolutionary inference depend heavily on an appropriate characterization of the underlying character substitution process.
In this paper, we present random-effects substitution models that extend common continuous-time Markov chain models into a richer class of processes capable of capturing a wider variety of substitution dynamics.
As these random-effects substitution models often require many more parameters than their usual counterparts, inference can be both statistically and computationally challenging.
Thus, we also propose an efficient approach to compute an approximation to the gradient of the data likelihood with respect to all unknown substitution model parameters.
We demonstrate that this approximate gradient enables scaling of sampling-based inference, namely Bayesian inference via Hamiltonian Monte Carlo, under random-effects substitution models across large trees and state-spaces.
Applied to a dataset of 583 SARS-CoV-2 sequences, an HKY model with random-effects shows strong signals of nonreversibility in the substitution process, and posterior predictive model checks clearly show that it is a more adequate model than a reversible model.
When analyzing the pattern of phylogeographic spread of 1441 influenza A virus (H3N2) sequences between 14 regions, a random-effects phylogeographic substitution model infers that air travel volume adequately predicts almost all dispersal rates.
A random-effects state-dependent substitution model reveals no evidence for an effect of arboreality on the swimming mode in the tree frog subfamily Hylinae.
Simulations reveal that random-effects substitution models can accommodate both negligible and radical departures from the underlying base substitution model.
We show that our gradient-based inference approach is over an order of magnitude more time efficient than conventional approaches.

\clearpage

\section{Introduction}

Along the branches of a phylogenetic tree, discrete characters such as nucleotides, amino acids, or morphologic traits evolve according to some (typically unknown) substitution process.
Substitution models are probabilistic representations of the substitution process and are central quantities in phylogenetic and phylodynamic models.
Broadly, substitution models describe the relative rates of discrete change from one character state to another.

When inferring phylogenies from character data, the nature of the substitution process is generally not the subject of primary biological interest.
Nevertheless, because substitution models stand as the key link between the phylogenetic tree and the observed discrete character data, appropriate modeling remains of paramount importance to avoid bias and their specification has received considerable attention \citep[see, \textit{e.g.},][]{tavare1986some,suchard2001bayesian,suchard2003inferring,woodhams2015new,abadi2019model,fabreti2022bayesian}.
There are also cases in which the substitution process is itself of direct interest.
In phylogeographic modeling of rapidly evolving pathogens, character states may represent the geographic locations of sampled pathogen sequences and the substitution process describes the spread of the pathogens through geographic space.
In this case, inferring an appropriately parameterized substitution model can deliver insight into the factors driving the spread of disease \citep{lemey2014unifying,dudas2017virus,lemey2020accommodating}.
Questions regarding potentially coevolving traits can also be addressed with substitution models by expanding their state-space, such as to pairs of binary characters \citep{pagel2006bayesian}.

Popular phylogenetic substitution models are continuous-time Markov chain (CTMC) models parameterized in terms of one or more infinitesimal rate matrices
and branch lengths that measure the expected number of substitutions along each branch in the phylogeny.
As the number of possible characters $\alphabetsize$ in the data state-space grows, the number of potential parameters in each rate matrix $\alphabetsize \times (\alphabetsize - 1)$ quickly becomes large.
When inferring a phylogeny from nucleotide sequences, the rate matrix is small, and relatively parameter-rich models have been considered \citep{tavare1986some,yang1994estimating}.
But rate matrices for other data types can easily grow large: there are 20 amino acids, 64 codons, and phylogeographic analyses can easily encompass many dozens of locations \citep{lemey2014unifying,dudas2017virus,gao2022new}.
Models which account for heterogeneity of the substitution process along branches, such as Markov-modulated models \citep{baele2021markov} may involve hundreds of parameters.
In such cases, inferring the unconstrained model, in which all $\order{S^2}$ non-diagonal elements are free parameters, has been historically prohibitive.
One reason is because the typical approach to Bayesian inference of substitution models is to use random-walk Metropolis-Hastings-based Markov chain Monte Carlo (MCMC) \citep{metropolis53,hastings1970monte}.
Such large rate matrices have many parameters which are (potentially) strongly correlated and often only weakly identifiable, rendering random-walk MCMC burdensome.

When confronted with substitution models for large state-spaces, the historical approach has been to find ways to reduce the number of free parameters in the model.
Amino acid models are often parameterized empirically \citep{dayhoff1978,whelan2001general}, requiring no free parameters for inference.
Codon models are often represented as combinations of site-level nucleotide models and codon-level processes \citep{yang2000codon}, some of which may be measured empirically \citep{hilton2018modeling}.
Such approaches reduce the number of parameters that must be inferred to $\bigO(\alphabetsize)$.
Another approach is to parameterize the rate matrix in terms of log-linear functions of observed covariates.
This generalized linear model (GLM) approach has been successful in phylogeographic inference, where observed covariates include factors like the distance between locations and air travel volumes \citep{lemey2014unifying,dudas2017virus}.
In addition to making inference tractable, the GLM approach can be used to quantify the strength of evidence for which factors do or do not affect the spread of infectious diseases.

In this paper, we demonstrate the utility of random-effects substitution models.
These models extend a wide class of CTMC models to incorporate additional rate variation by representing the original (base) model as fixed-effect model parameters and allowing the additional random-effects to capture deviations from the simpler process.
We demonstrate the utility of these models in Bayesian inference on a variety of exemplar evolutionary problems.
On a dataset of 583 SARS-CoV-2 genomes, an HKY model with random-effects captures known mutational biases in SARS-CoV-2 and is shown to be superior to the richest, general-time reversible (GTR) model.
Applied to a phylogeographic analysis of influenza A subtype H3N2, a GLM substitution model with random-effects provides evidence that air travel volume captures the geographic process of dispersal for all except a small set of pairs for which it underpredicts dispersal.
As a test for ecologically-dependent trait evolution, a random-effects pairwise-dependent substitution model finds no evidence for an effect of arboreality on the swimming mode in hylid tree frogs.
We quantify the performance of random-effects substitution models using simulations.

Before we can perform inference under these models, however, there stand two obstacles that must be overcome.
First and foremost, the parameter-space of the random effects models can be very large, and these parameters may be strongly correlated.
To overcome the dimensionality, we derive an efficient-to-compute approximation of the gradient of the phylogenetic (log-)likelihood with respect to (wrt) all of the parameters simultaneously of an arbitrary CTMC substitution model.
Notably the exact gradient is often computationally prohibitive.
We implement our approximate gradient in the phylogenetic inference software package BEAST 1.10 \citep{suchard2018bayesian} and the high-performance computational library BEAGLE 3 \citep{ayres2019beagle}, enabling the use of Hamiltonian Monte Carlo (HMC), a gradient-based alternative to random-walk MCMC \citep{neal2011mcmc}, for efficient parameter inference.
HMC leverages gradients to take bold steps through even highly correlated parameter spaces and can greatly increase MCMC efficiency.
Second, to avoid identifiability issues with potentially overparameterized models (such as inferring a $14 \times 14$ rate matrix based on a single observed character), we make use of the Bayesian bridge prior \citep{polson2014bayesian} that is strongly regularizing and allows the data to decide which parameters are important to capture their variability.

The rest of this paper is structured as follows.
In the Methods section, we formally introduce the random-effects substitution model and the Bayesian bridge prior distribution.
Then we derive our approximate gradient of the phylogenetic log-likelihood with respect to parameters of the substitution model.
We also provide an introduction to gradient-based inference.
In the Results section, we first investigate the increase in efficiency from using our approximate gradients compared to alternative approaches, both for optimization tasks and full Bayesian inference.
Then we apply our random-effects substitution model to a number of real-world examples and to simulated data.
We conclude by contemplating future approaches for improving inference efficiency and additional application areas where random-effects substitution models are likely to be useful.

\section{Methods}
In this paper we assume that there is a (possibly unknown) rooted phylogeny $\phylogeny$ with $\nTaxa$ tips that links the observed character sites and $\nTaxa - 1$ internal nodes.
We index the branch lengths and nodes such that the edge connecting node $\node$ to its parent $\parent$ has length $\branchlength_{\node}$.
Along each branch of the tree, we assume that characters arise from an alphabet of size $\alphabetsize$ and evolve under a CTMC model with instantaneous rate matrix $\ratematrix = \{ \ratematrixelement_{ij} \}$, where $\ratematrixelement_{ij} \ge 0$ for $i \neq j$ and the diagonal elements are fixed such that row-sums of $\ratematrix$ equal 0.
We measure branch length $\branchlength_{\node}$ in expected number of substitutions per site according to a probability mass vector $\boldsymbol{\pi} = \left(\pi_1,\ldots,\pi_{\alphabetsize} \right)$ over the characters.
Often, $\boldsymbol{\pi}$ is taken as the stationary distribution of $\ratematrix$, but this need not be the case.
To account for variation in evolutionary rates across sites in the character alignment, finite mixture models \citep[\textit{e.g.}, the discrete-gamma model of][]{yang1994maximum} modulate the expected number of substitutions along all branches at a specific site.
Consider that there are $\nCategories$ rate categories, the rate scalar in the $r$th category is $\rateCategory_\category$, and the prior probability of being in any particular mixture category is $\prob{\rateCategory_{\category}}$.
Then, the finite-time transition probability matrix along branch $\node$ in category $\category$ is given by $\bP_{\node \category} = \exp(\rateCategory_{\category} \times \ratematrix \times \branchlength_{\node} )$, where we assume that $\ratematrix$ is normalized wrt $\bpi$.
The matrix $\bP_{\node \category}$ governs the probability of change from state $i$ to state $j$ along branch $\node$ in category $\category$.
Note that the subscripts on $\bP$ do not denote elements of the matrix but rather which of the $\nCategories \times (2\nTaxa - 2)$ distinct transition probability matrices---one for each branch and rate category---is under consideration.
In truth, the rate matrix $\ratematrix$ is a function of a vector of estimable parameters $\params$, specifically $\ratematrix(\params)$, but we suppress this notation for ease of presentation.

\subsection{Random-effects substitution models}
Random-effects substitution models are extensions of simpler CTMC substitution models.
We start with a \emph{base model}, which could be as simple as Jukes-Cantor \citep{jukes1969evolution}, as complex as a GLM substitution model \citep{lemey2014unifying} or anything in between.
This base model carries a rate matrix $\mathbf{B} = \{ b_{ij} \}$ and probability mass vector $\boldsymbol{\pi}_{\mathbf{B}}$ over the characters.
We define the random-effects substitution model rate matrix $\ratematrix$ using the following log-linear formulation,
\begin{equation}
\log \ratematrixelement_{ij} = \log b_{ij} + \randomeffect_{ij}
  \text{ for } i \neq j ,
  \label{eqn:refx_model}
\end{equation}
and set $\boldsymbol{\pi} = \boldsymbol{\pi}_{\mathbf{B}}$.
Intuitively, the random-effects $\randomeffect_{ij}$ are multiplicative real-valued parameters which enable each non-diagonal element to deviate from the values specified by the base model.
For example, $\randomeffect_{ij} = \log(2)$ doubles the rate implied by the base model, $\ratematrixelement_{ij} = 2 b_{ij}$.

Random-effects substitution models retain the basic structure of the base model that may be biologically or epidemiologically motivated, while allowing for potentially large deviations from this base process.
In our phylogeographic example, we start with an epidemiologically motivated model for the spread of rapidly evolving pathogens parameterized using air travel volume between countries with a GLM substitution model (though we could potentially use many more covariates) and allow the random-effects to capture shortcomings of this description.
For convenience, we shorthand the random-effects version of a substitution model $\basemodel + \text{RE}$, where $\basemodel$ is the base model (e.g.~HKY).

\subsection{Bayesian regularization}
Random-effects substitution models are in general overparameterized and as such not identifiable by the likelihood alone.
In a Bayesian setting, a prior will provide relief from this and allow for a posterior to be inferred.
Nevertheless, such circumstances demand careful thought about the choice of prior, as it will play a key role in determining the posterior.
An attractive class of priors for these situations are shrinkage priors such as the Bayesian bridge \citep{polson2014bayesian} or the horsehoe \citep{carvalho2010horseshoe}.
Shrinkage priors were originally developed for handling regression models when there are more parameters than observations.
The priors induce sparsity in the model by strongly regularizing coefficients to be near 0 when the data provide little or no information to the contrary and, otherwise, impart little bias into the posterior.
Shrinkage priors have also found success in phylogenetic contexts, including models for the rate of evolution of the rate of evolution \citep{fisher2021shrinkage,fisher2021relaxed}, population sizes over time \citep{faulkner2020horseshoe}, and rates of speciation and extinction \citep{magee2020locally}.
In these cases, the prior pull is towards no change, either from a branch to its descendants or from one time interval to the next.
In the context of random-effects substitution models, sparsity is imposed by pulling the random-effects towards the null value of 0 (at 0 there is no deviation from the base model in the $i \rightarrow j$ direction and $\lambda_{ij} = b_{ij}$).

Shrinkage priors also permit us to perform model selection, with the data and prior reconciling which parameters belong in or out of the model.
For the random-effects substitution model, a particular random-effect $\randomeffect_{ij}$ is excluded from the model if it is approximately 0.
Where discrete mixture models, such as those used in Bayesian stochastic search variable selection \citep[BSSVS,][]{lemey2009bayesian}, carry a finite probability that a parameter achieves exactly 0, shrinkage priors instead have a large spike of prior density near 0.
While this occasionally makes it more difficult to declare if a parameter belongs in the model or not, the use of  purely continuous priors usually yields Markov chains that mix more efficiently and, importantly, permits the use of gradient-based inference.

The Bayesian bridge prior on random-effect $\randomeffect$ has density,
\begin{equation}
  \pr(\randomeffect \mid \globalScale, \bridgeExponent) \propto \text{exp} \left( - \left| \frac{\randomeffect}{\globalScale} \right|^{\bridgeExponent} \right),
\end{equation}
where the global scale $\globalScale$ controls the overall spread of the distribution and the exponent $\bridgeExponent$ controls the shape.
The Bayesian bridge is perhaps best thought of as family of distributions, modulated by $\bridgeExponent$, some of which fall into the class of shrinkage priors ($\alpha < 1$) and some of which do not.
At $\bridgeExponent = 2$, the density coincides with a standard normal distribution, while at $\bridgeExponent = 1$ it is the density of the Laplace distribution.
At lower exponent values, the distribution becomes increasingly peaked around 0 and induces sparsity.
We use $\bridgeExponent = 1/4$, which in practice imposes a useful level of sparsity without compromising MCMC convergence.
The global scale $\globalScale$ also plays an important role in determining the degree of regularization.
To both permit the data to inform the strength of regularization and efficient Gibbs sampling procedures, we place a Gamma(shape=1, rate=2) prior on $\globalScale^{-\bridgeExponent}$ \citep{nishimura2023shrinkage}.

The Bayesian bridge distribution has particularly fat tails for lower $\bridgeExponent$.
This can hamper sampling, and it can allow parameter values which are, \textit{a priori}, unrealistically large (or small).
Particularly large random-effects can also cause numerical instability when exponentiating the substitution-rate matrix.
These effects can be ameliorated by the use of the shrunken-shoulder Bayesian bridge \citep{nishimura2023shrinkage}.
This formulation includes a ``slab'' parameter $\zeta$ that controls the tails of the distribution.
Specifically, outside of $[-\zeta,\zeta]$, the tails of the shrunken-shoulder Bayesian bridge become Normal(0,$\zeta^2$).
We set $\zeta = 2$, which \textit{a priori} specifies that it is unlikely for a particular element of the rate matrix to be more than $e^2 \approx 7$ times larger or smaller than specified by the base model.

\subsection{The gradient of the phylogenetic log-likelihood}

In this paper we are interested in the gradient of the phylogenetic log-likelihood with respect to the parameters of the substitution model.
The data $\data$ are a collection of homologous sites (columns in a multiple sequence alignment), $\data = (\data_1, \dots, \data_\nSites)$.
We will write the likelihood $\likelihoodAll{\params}$, and its gradient $\nabla \likelihoodAll{\params}$.
The gradient is the collection of derivatives wrt all substitution model parameters,
\begin{equation}
	\nabla \likelihoodAll{\params} = \left( \frac{\partial}{\partial \param_1} \likelihoodAll{\params}, \dots, \dBydParam \likelihoodAll{\params} \ \right)\transpose ,
\end{equation}
where $\transpose$ denotes the transpose operator.

Under the common assumption that sites evolve independently and identically, we can express the log-likelihood as a sum across all $C$ sites, and hence derivatives of it as well.
We have
\begin{align}
	\dBydParam \log \likelihoodAll{\params} &= \sum_{\site=1}^{\nSites} \dBydParam \log \left( \sum_{\category=1}^\nCategories \likelihood{\params, \category} \prob{\rateCategory_{\category}} \right) \nonumber \\
	&= \sum_{\site=1}^{\nSites} \frac{\sum_{\category=1}^\nCategories \dBydParam \likelihood{\params, \category} \prob{\rateCategory_{\category}}}{\sum_{\category=1}^\nCategories \likelihood{\params, \category} \prob{\rateCategory_{\category}}}
\end{align}
The denominator is simply the likelihood of a site $\likelihood{\params}$.
For simplicity, we will focus on the computation of $\dBydParam \likelihood{\params, \category}$ for site $\site$ under rate category $\category$.

\subsubsection{Partial likelihood vectors and the phylogenetic likelihood}
We can, at any node $\node$ in the tree, compute the likelihood as
\begin{equation}
	\likelihood{\params} =
	 \sum_{r=1}^R  \left[ \bpostPartial_{\node \category \site}\transpose \bprePartial_{\node \category \site} \right] \prob{\rateCategory_{\category}}.
	 \label{eqn:likelihood_from_partials}
\end{equation}
The post-order partial likelihood vector $\bpostPartial_{\node \category \site}$, describes the probability, at node $\node$, in rate category $\category$, at the $\site$th site, of observing the tip-states in all tips which descend from the node, conditioned on the state at the node.
The pre-order partial likelihood vector, $\bprePartial_{\node \category \site}$, describes the \textit{joint} probability of observing the tip-states in all tips \textit{not} descended from the node and the state at the node.

\begin{figure}
	\centering
	\includegraphics[width=0.4\textwidth]{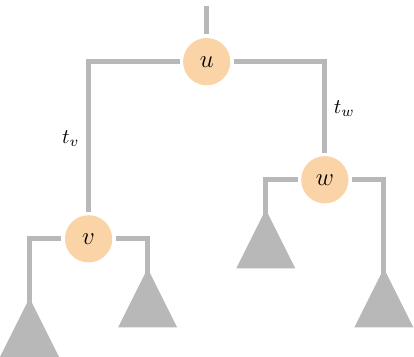}
	\caption{
	A phylogenetic tree highlighting three key nodes.
	We will take node $\node$ as our focal node, which here has parent $\parent$ and sister $\sister$.
	We index branch lengths by the node which subtends them, such that the branch with length $\branchlength_{\node}$ is the branch leading to node $\node$.
	}%
	\label{fig:tree_with_labels}
\end{figure}

The post-order partial likelihood vectors are computed via pruning from the tip to the roots (a post-order traversal), for the tree in Figure~\ref{fig:tree_with_labels}, via
\begin{align}
	\bpostPartial_{\parent \category \site} = \bP_{\node \category} \bpostPartial_{\node \category \site} \circ \bP_{\sister \category} \bpostPartial_{\sister \category \site}.
\end{align}
The pre-order partial likelihood vectors are then computed in a root-to-tip pass through the tree (a pre-order traversal) using the relation
\begin{align}
	\bprePartial_{\node \category \site} = \bP_{\node \category}\transpose [\bprePartial_{\parent \category \site} \circ \bP_{\sister \category} \bpostPartial_{\sister \category \site}].
	\label{eqn:preorder}
\end{align}
We note that $\bpostPartial_{\node \category \site}$ is independent of $\bP_{\node \category}$, while $\bprePartial_{\node \category \site}$ is dependent on $\bP_{\node \category}$.
At the root $\node = 2 \nTaxa - 1$, the pre-order partial likelihood vector is simply the root-frequency vector $\rootFreqs$, which may or may not be the same as the frequency vector $\boldsymbol{\pi}$ used to normalize the rate matrix.

\subsubsection{A na\"ive derivative}
We can use the multivariable chain rule to obtain the total derivative of the likelihood wrt $\param_k$.
To do this, we first envision a simple parameter expansion with branch- and root-specific variables $\perBranchParam_{\node k} = \theta_k$ and rewrite the differential as
\begin{equation}
	\frac{\partial}{\partial \param_k} 
	= \sum_{\node = 1}^{2 \nTaxa - 1} 
	\left( \frac{\partial}{\partial \perBranchParam_{\node k}} \right)
	\left( \frac{\partial \perBranchParam_{\node k}}{\partial \theta_k} \right)
	= \sum_{\node = 1}^{2 \nTaxa - 1} 
	\frac{\partial}{\partial \perBranchParam_{\node k}} .
\end{equation}
Then, the partial likelihood vector representation of the phylogenetic likelihood allows us to isolate the contribution of each branch and $\rootFreqs$ to this total derivative.
In doing so, we also recall that $\bpostPartial_{\node \category \site}$ is independent of $\bP_{\node \category}$ and $\rootFreqs$, such that $\left( \partial / \partial  \perBranchParam_{\node k} \right) \bpostPartial_{\node \category \site} = 0$ for all $v$. 
By summing over all branches and the root, we obtain the total derivative as 
\begin{align}
	\frac{\partial}{\partial \param_k} \likelihood{\params, \category} &= \sum_{\node = 1}^{2 \nTaxa - 1} \dBydPerBranchParam \bpostPartial_{\node \category \site}' \bprePartial_{\node \category \site} \nonumber \\
  &= \sum_{\node = 1}^{2 \nTaxa - 1} \bpostPartial_{\node \category \site}' \dBydPerBranchParam \bprePartial_{\node \category \site} \nonumber \\
  &= \sum_{\node = 1}^{2 \nTaxa - 2} \bpostPartial_{\node \category \site}' \left( \dBydPerBranchParam \bP_{\node \category} \right)' [\bprePartial_{\parent} \circ \bP_{\sister \category} \bpostPartial_{\sister}] + \rootLabel \nonumber \\
  &= \sum_{\node = 1}^{2 \nTaxa - 2} \bpostPartial_{\node \category \site}' \left( \dBydPerBranchParam \bP_{\node \category} \right)' \bprePrePartial_{\node \category \site}  + 
  \rootLabel
  \nonumber \\
  &= \sum_{\node = 1}^{2 \nTaxa - 2} \bpostPartial_{\node \category \site}' \left( \frac{\partial}{\partial \perBranchParam_{\node k}} e^{\ratematrix \times t_{\node}  \times \rateCategory_{\category}} \right)' \bprePrePartial_{\node \category \site} + \rootLabel \nonumber \\
  &= \sum_{\node = 1}^{2 \nTaxa - 2} \bpostPartial_{\node \category \site}' \left( \sum_{i=1}^{\alphabetsize} \sum_{j=1}^{\alphabetsize} \frac{\partial e^{\ratematrix \times t_{\node} \times \rateCategory_{\category}}}{\partial \ratematrixelement_{ij}} \frac{\partial \ratematrixelement_{ij}}{\partial \perBranchParam_{\node k}} \right)' \bprePrePartial_{\node \category \site} + \rootLabel,
  \label{eqn:naive_phylo_gradient}
\end{align}
where the contribution from root-frequency vector is
\begin{equation}
\rootLabel = \rootContribution .
\end{equation}
In the third-to-last step, we defined $\bprePrePartial_{\node \category \site} = [\bprePartial_{\parent} \circ \bP^{(\sister)} \bpostPartial_{\sister}]$ to simplify the notation and focus on the part of the equation which depends on $\perBranchParam_{\node k}$ ($\param_k$).
In the last step, we employed the matrix chain rule \citep{petersen2008matrix}.
The term $\partial \exp(\ratematrix \times t_{\node} \times \rateCategory_{\category})/\partial \ratematrixelement_{ij}$ is the derivative of the matrix exponential with respect to one of the elements of the rate matrix, which we discuss in more detail in the Section ``Efficiently approximating the derivative of the matrix exponential.''
We note that the rate matrix $\ratematrix$ is sometimes defined to be unnormalized, such that the transition probability matrix along a branch is instead given by $\bP_{\node \category} = \exp(\rateCategory_{\category} \times (-1/(\operatorname{diag}(\ratematrix)' \, \boldsymbol{\pi})) \times \ratematrix \times \branchlength_{\node} )$.
For simplicity of notation, when normalization is desired we take the rate matrix $\ratematrix$ to be normalized, and allow the $\partial \ratematrixelement_{ij} / \partial \perBranchParam_{\node k}$ term to capture the effect of the normalizing constant on the elements of the rate matrix.

As we discuss in the Supplemental Section ``Computational complexity of alternative approaches to computing the gradient of a matrix exponential,'' the computational cost of obtaining $\partial \exp(\ratematrix \times t_{\node} \times \rateCategory_{\category})/\partial \ratematrixelement_{ij}$ is $\bigO(\alphabetsize^3)$.
The sum in Equation~\ref{eqn:naive_phylo_gradient} requires this quantity for all $\alphabetsize^2$ elements in $\ratematrix$ and for each of the $2 \nTaxa - 2$ branches, making the cost to compute the derivative $\bigO(\nTaxa \alphabetsize^5)$.
Obtaining the gradient requires using Equation~\ref{eqn:naive_phylo_gradient} for all $K$ substitution model parameters, making the cost of the gradient $\bigO(K \nTaxa \alphabetsize^5)$.
For random effects models, this is $\bigO(\nTaxa \alphabetsize^7)$.
Such costs are prohibitive for even moderate $\alphabetsize$, so we turn our attention now to improving the computational efficiency of gradient computations.

\subsubsection{Reducing the computational complexity}\label{sec:mapping}

We can reformulate the na\"ive approach of Equation~\ref{eqn:naive_phylo_gradient} to produce a more efficient gradient computation.
By rearranging the order of summation, we can disentangle the derivative of the rate matrix wrt its elements from the derivative of its elements wrt model parameters.
Specifically,
\begin{align}
	\dBydParam \likelihood{\params, \category} &= \sum_{\node = 1}^{2 \nTaxa - 2} \bpostPartial_{\node \category \site}' \left( \sum_{i=1}^{\alphabetsize} \sum_{j=1}^{\alphabetsize} \frac{\partial e^{\ratematrix \times t_{\node} \times \rateCategory_{\category}}}{\partial \ratematrixelement_{ij}} \frac{\partial \ratematrixelement_{ij}}{\partial \perBranchParam_{\node k}} \right)' \bprePrePartial_{\node \category \site} \nonumber \\
	&= \sum_{\node = 1}^{2 \nTaxa - 2} \sum_{i=1}^{\alphabetsize} \sum_{j=1}^{\alphabetsize} \bpostPartial_{\node \category \site}' \left( \frac{\partial e^{\ratematrix \times t_{\node} \times \rateCategory_{\category}}}{\partial \ratematrixelement_{ij}} \right)' \bprePrePartial_{\node \category \site} \frac{\partial \ratematrixelement_{ij}}{\partial \perBranchParam_{\node k}} \nonumber \\
	&= \sum_{\node = 1}^{2 \nTaxa - 2} \mapping_{k} \vecm(\allQderivs_{\node}),
	\label{eqn:almost_the_full_mapping}
\end{align}
where the $\vecm$ operator makes the matrix $\allQderivs_{\node}$ into a column vector by stacking the columns on top of each other, and we obtain the last line by defining two new quantities which we will now discuss.

The matrix $\mapping$ is a $K \times \alphabetsize^2$ mapping matrix, which stores in each row $k$ a vector of the partial derivatives of all elements of $\ratematrix$ wrt $\param_k$,
\begin{equation}
	\mapping_{k \boldsymbol{\cdot}} = \left( \frac{ \partial \ratematrixelement_{11} }{ \partial \param_k}, \dots, \frac{ \partial \ratematrixelement_{\alphabetsize \alphabetsize} }{ \partial \param_k} \right).
\end{equation}
The matrix $\allQderivs_{\node \category} = \{ \allQderivselement_{\node \category i j} \}$ contains the contribution of branch $\node$ to the derivative of the phylogenetic likelihood wrt the $ij^{\text{\tiny th}}$ entry of $\ratematrix$ in rate category $\category$.
Specifically,
\begin{equation}
  \allQderivselement_{\node \category i j} = \bpostPartial_{\node \category \site}' \left( \frac{\partial}{\partial \ratematrixelement_{ij}} e^{\ratematrix \times t_{\node} \times \rateCategory_{\category}} \right)' \bprePrePartial_{\node \category \site} .
  \label{eqn:all_q_derivs_element}
\end{equation}

We arrive at the entire gradient (as opposed to a single entry) and increase computational efficiency by replacing $\mapping_{k \boldsymbol{\cdot}}$ with $\mapping$ in Equation~\ref{eqn:almost_the_full_mapping} and rearranging,
\begin{align}
	\nabla \likelihood{\params, \category} &= \sum_{\node} \mapping \vecm(\allQderivs_{\node}) \nonumber \\
	&= \mapping \sum_{\node} \vecm(\allQderivs_{\node}).
	\label{eqn:derivs_with_mappings}
\end{align}

This approach separates the gradient of the phylogenetic likelihood wrt model parameters into two pieces, a gradient of the phylogenetic likelihood wrt elements of the rate matrix, and a gradient of the elements of the rate matrix wrt the model parameters.
The result is the intermediate quantity $\allQderivs_{\node}$ that can be obtained with only a single computation of the derivative of a matrix exponential per branch.
As this quantity can be summed across the tree prior to mapping it to the substitution model parameters, $\bigO(\nTaxa)$ matrix multiplications are avoided.
The result is that this approach is $\bigO(K \alphabetsize^2 + \nTaxa \alphabetsize^5)$ rather than $\bigO(K \nTaxa \alphabetsize^5)$.
For random-effects substitution models, this is the difference between a $\bigO(\nTaxa \alphabetsize^7)$ computation and a $\bigO(\alphabetsize^4 + \nTaxa \alphabetsize^5) = \bigO(\nTaxa \alphabetsize^5)$ computation.
Note that this approach works for branch-specific models as well, by specifying the mapping matrix $\mapping$ appropriately.

\subsubsection{Efficiently approximating the derivative of the matrix exponential}\label{sec:approximation}
We now turn our attention to an efficient approximation to the derivative of a matrix exponential.
The derivative of a matrix exponential can be represented as a power-series \citep[][Equation 103]{najfeld1995derivatives},
\begin{equation}
	\frac{\partial}{ \partial \ratematrixelement_{ij}} e^{\ratematrix \branchlength} = e^{\ratematrix \branchlength} \sum_{x=0}^{\infty} \frac{\branchlength^{x+1}}{(x+1)!} \{\indicatormatrixij,\ratematrix^x\},
	\label{eqn:NH_103}
\end{equation}
where $\indicatormatrixij$ is a matrix which is 0 for all but the $(ij)$th entry, which is 1.
The matrix commutator power $\{\indicatormatrixij,\ratematrix^x\}$ for non-negative integer $x$ is defined recursively \citep{najfeld1995derivatives}, such that $\{\indicatormatrixij,\ratematrix^0\} = \indicatormatrixij$ and $\{\indicatormatrixij,\ratematrix^x\} = [\{\indicatormatrixij,\ratematrix^{x-1}\},\ratematrix]$ (where $[\bA,\bB]$ is the matrix commutator $\bA \bB - \bB \bA$).

The first-order approximation to Equation~\ref{eqn:NH_103} is taken by keeping only the $x=0$ term, yielding
\begin{equation}
	\frac{\partial}{ \partial \ratematrixelement_{ij}} e^{\ratematrix \branchlength} \approx \branchlength e^{\ratematrix \branchlength} \indicatormatrixij.
	\label{eqn:exponential_approximate_gradient}
\end{equation}
We can use this first-order approximation to approximate $\allQderivs_{\node}$ on each branch.
Specifically,
\begin{align}
	\allQderivselement_{\node i j} &\approx \bpostPartial_{\node \category \site}' [(t_{\node} \times \rateCategory_{\category} \times e^{\ratematrix \times t_{\node} \times \rateCategory_{\category}} \indicatormatrixij)' [\bprePartial_\parent \circ (e^{\ratematrix \branchlength_{\sister} \times \rateCategory_{\category}} \bpostPartial_{\sister})]] \nonumber \\
	&= t_{\node} \times \rateCategory_{\category} \times \bpostPartial_{\node \category \site}' [ \indicatormatrix_{ji} (e^{\ratematrix \times t_{\node} \times \rateCategory_{\category}})' [\bprePartial_\parent \circ (e^{\ratematrix \times t_{\node} \times \rateCategory_{\category}} \bpostPartial_{\sister})]] \nonumber \\
	&= t_{\node} \times \rateCategory_{\category} \times \bpostPartial_{\node \category \site}' \indicatormatrix_{ji} \bprePartial_{\node \category \site} \nonumber \\
	&= t_{\node} \times \rateCategory_{\category} \times \prePartial_{\node i} \postPartial_{\node j} ,
	\label{eqn:approximate_gradient}
\end{align}
where we get from line 3 to line 4 by noting that $(e^{\ratematrix \times t_{\node} \times \rateCategory_{\category}})' = \bP'_{\node \category}$ and applying Equation~\ref{eqn:preorder}.
Intuitively, we have the (approximate) derivative with respect to an $i \rightarrow j$ transition depending on the pre-order partial likelihood in state $i$ and the post-order partial likelihood in state $j$.

Equation~\ref{eqn:approximate_gradient} means that we can write our approximate $\allQderivs_{\node}$ as an outer product,
\begin{align}
  \allQderivs_{\node \category \site} \approx \branchlength_{\node} \times \rateCategory_{\category} \times \bprePartial_{\node \category \site} \otimes \bpostPartial_{\node \category \site}.
  \label{eqn:approximate_gradient_outer_product}
\end{align}
This means that we can obtain all $\alphabetsize^2$ entries of $\allQderivs_{\node \category \site}$ in $\bigO(\alphabetsize^2)$, which is much more efficient than the $\bigO(\alphabetsize^5)$ cost of the non-approximate computation.
Thus, with this approximation and the mapping approach outlined in the previous section, the (approximate) substitution gradient can be obtained in $\bigO(K \alphabetsize^2 + \nTaxa \alphabetsize^3)$, rather than the $\bigO(K \alphabetsize^2 + \nTaxa \alphabetsize^5)$ cost suggested by Equation~\ref{eqn:derivs_with_mappings} or the $\bigO(K \nTaxa \alphabetsize^5)$ cost suggested by Equation~\ref{eqn:naive_phylo_gradient}.
We will denote the approximate gradient that comes from using this approximation to $\allQderivs_{\node \category \site}$ in Equation~\ref{eqn:derivs_with_mappings} as $\approxnabla \log \likelihoodAll{\params}$.

\subsection{Hamiltonian Monte Carlo with surrogate trajectories}\label{sec:inference}

HMC \citep{duane1987hybrid,neal2011mcmc} is an advanced MCMC algorithm that falls broadly within the well-known class of Metropolis-Hastings MCMC (MH-MCMC) algorithms \citep{metropolis53,hastings1970monte}.
By allowing samples to be drawn (sequentially) from arbitrary target distributions, MH-MCMC algorithms like HMC allow users to approximate distributions that do not have known closed-form densities.
Unlike many commonly-employed random-walk Metropolis-Hastings proposals, however, HMC uses information captured by the log-posterior gradient to traverse a model's parameter space much more efficiently by proposing (and accepting) states which are farther apart.

HMC constructs an artificial Hamiltonian system by augmenting the parameter space to include an auxiliary Gaussian `momentum' variable $\mom \sim$ MVN$_K(\zero,\cov)$ that is independent from the target variable $\params$ by construction.
Letting $\posteriorFunction(\cdot)$ denote the posterior density, the resulting Hamiltonian energy function is the negative logarithm of the joint distribution over $\params$ and $\mom$.
Ignoring normalizing constants we obtain
\begin{align*}
	H(\params,\mom) \,=\, - \log \posteriorFunction(\params)  + \frac{1}{2} \mom^T \cov^{-1} \mom \, ,
\end{align*}
and Hamilton's equations are
\begin{align*}
	\dot{\params} &= + \frac{\partial H}{\partial \mom} = \cov^{-1} \mom \\ \nonumber
	\dot{\mom}    &= - \frac{\partial H}{\partial \params} =  \nabla \log \posteriorFunction(\params)  \, .
\end{align*}
On the one hand, one may show that the action of the dynamical system that satisfies these equations leaves the target $\posteriorFunction(\cdot)$ invariant thanks to the reversibility, volume preservation and energy conservation of Hamiltonian dynamics.
On the other hand, closed-form descriptions of these dynamics are rarely available for arbitrary target distributions, leading to the need for computer intensive approximations.
In particular, the St\"{o}rmer-Verlet (velocity Verlet) or leapfrog method \citep{leimkuhler2004simulating} has become the numerical integrator of choice for obtaining discretized trajectories within HMC.
Beginning at time $\tau$ and letting $\epsilon>0$ be small, a single leapfrog iteration proceeds thus:
    \begin{align}\label{eq:leap}
	\mom\left(\tau + \frac{\epsilon}{2}\right) &= \mom(\tau) + \frac{\epsilon}{2} \nabla \log \posteriorFunction(\params(\tau))  \\  \nonumber
	\params(\tau+\epsilon) &= \params(\tau) + \epsilon \, \cov^{-1}\mom(\tau+\frac{\epsilon}{2}) \\ \nonumber
	\mom(\tau + \epsilon) &= \mom\left(\tau+\frac{\epsilon}{2}\right) + \frac{\epsilon}{2} \nabla \log \posteriorFunction(\params(\tau+\epsilon))  \, .
\end{align}
Trajectories arising from concatenated leapfrog iterations maintain some of the desirable qualities of the exact Hamiltonian dynamics (reversibility, voume preservation) but no longer conserve energy.
For this reason, the HMC algorithm features three distinct steps.
First, it draws an initial momentum.
Then, it uses that momentum and numerically-integrated dynamics to generate a proposal for a new value of $\params$.
Lastly, it either accepts or rejects this new value according to the usual Metropolis-Hastings acceptance probability rule \citep{metropolis53,hastings1970monte}.
The accept/reject step accounts for integration error and leaves the target distribution invariant.

Indeed, HMC's Metropolis correction allows for additional deviations from Hamiltonian dynamics over and beyond numerical discretization schemes such as \eqref{eq:leap}.
\emph{Surrogate HMC} methods seek to improve computational performance of HMC by approximating the log-posterior gradient with less expensive surrogate models including, e.g., piecewise-approximations \citep{zhang2017precomputing}, Gaussian processes \citep{rasmussen2003gaussian,lan2016emulation} or neural networks \citep{zhang2017hamiltonian,li2019neural}.
Directly relevant to the present work, \citet{li2019neural} show the validity of replacing the log-posterior gradient $\nabla \log \posteriorFunction(\cdot)$ within the leapfrog method \eqref{eq:leap} with \emph{any} vector function $\mathbf{g}: \mathbb{R}^K \rightarrow \mathbb{R}^K$.
In particular, such an approach maintains the reversibility and volume preservation of Hamiltonian dynamics and, when paired with Metropolis corrections, leaves the target posterior distribution invariant.
In the present work, we select $\mathbf{g} = \approxnabla\log \likelihoodAll{\params} + \nabla \log \prob{\params}$, the approximate posterior gradient obtained by using our approximation to the gradient of the phylogenetic log-likelihood and the true gradient for the prior.
In the supplementary materials, we discuss an alternative justification.

\section{Results}

\subsection{C to T bias in SARS-CoV-2 evolution}
\label{sec:SC2}

The mutational profile of SARS-CoV-2 has been intensely scrutinized, one feature in particular which has been noted is a strongly increased rate of C$\rightarrow$T substitutions over the reverse T$\rightarrow$C substitutions.
We note that while RNA viruses like SARS-CoV-2 use uracil (U) in place of thymine (T), it is generally coded as thymine--the coding of adenosine (A), cytosine (C), and guanine (G) are unchanged.
The elevation of one direction of substitution over its reverse is a violation of the common phylogenetic assumption of reversibility made by the GTR \citep{tavare1986some} family of substitution models.
Random-effects substitution models are suitable for addressing this model violation, in particular we consider an HKY+RE substitution model.
In principle we could choose any GTR-family model.
HKY represents a balance between the simplicity of JC+RE (where the random-effects would also have to account for uneven nucleotide frequencies) and the complexity of GTR+RE (where the random-effects only capture nonreversibilities).
The rate matrix is
\begin{equation}
	\log \ratematrixelement_{ij} =  \log \kappa \times \indicator(ij \in \mathcal{T}) + \log \pi_j + \randomeffect_{ij},
\end{equation}
where $\kappa$ is the HKY parameter governing relative rate of transitions to transversions, $\indicator(ij \in \mathcal{T})$ indicates that the $i$ to $j$ change is a transition, and $\bpi$ are the HKY stationary frequencies.

\begin{figure}[h]
	\centering
	\includegraphics[width=0.85\textwidth]{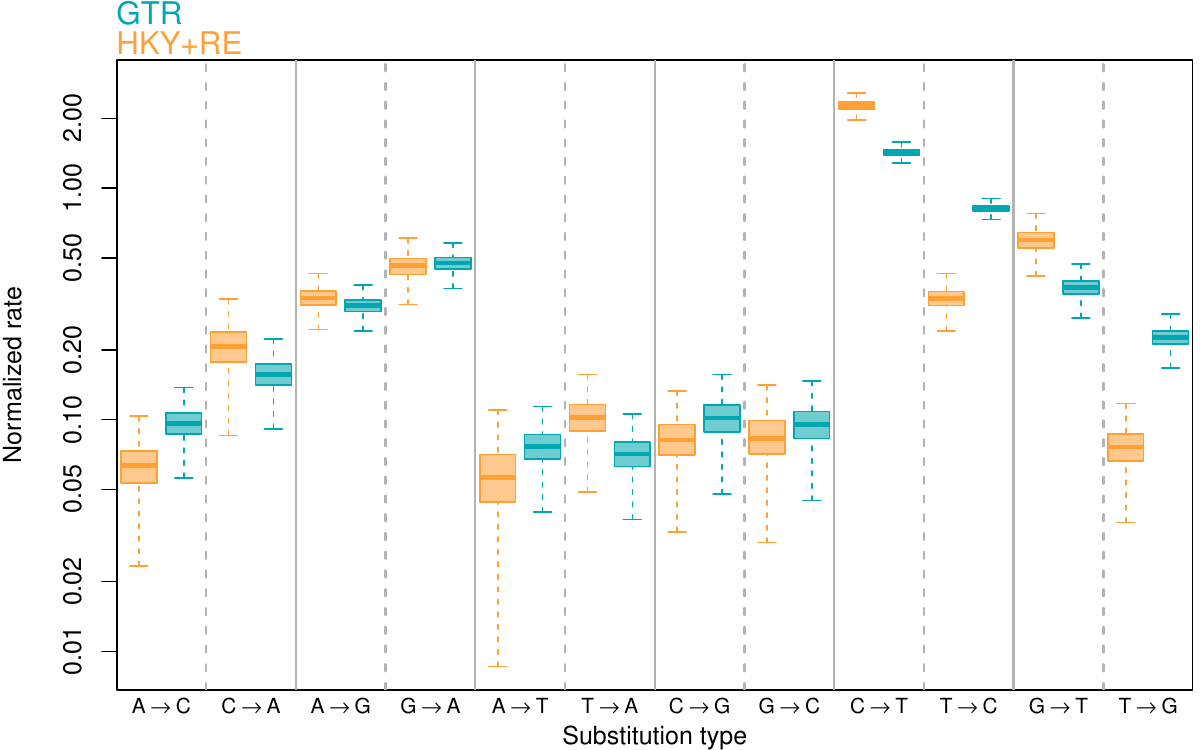}
	\caption{
	Posterior distributions of the 12 non-diagonal elements of the inferred rate matrices for the dataset of \citet{pekar2021timing}.
	The solid line is the posterior median, the shaded region the 50\% CI.
	The whiskers extend to the posterior samples farthest from the median but within 1.5$\times$ the interquartile range.
  	By comparing HKY+RE (which is not constrained by the assumption of reversibility) to GTR, we can see that the assumption of reversibility leads to the overestimation of the T$\rightarrow$C (and T$\rightarrow$G) rates and the underestimation of the C$\rightarrow$T (and G$\rightarrow$T) rates.
	}%
	\label{fig:refx_vs_gtr_rates}
\end{figure}

We apply this HKY+RE model to infer both the dynamics of molecular substitution and the phylogeny for 583 SARS-CoV-2 sequences from \citet{pekar2021timing}.
(More information about the model and dataset is in Supplemental Table S1.)
Consistent with previous studies \citep[\textit{e.g.}][]{matyavsek2020mutation,tonkin2021patterns}, we find evidence for a greatly elevated rate of C$\rightarrow$T substitutions, as well as an elevated G$\rightarrow$T rate (Figure~\ref{fig:refx_vs_gtr_rates}).
We can test the support for nonreversibilities, for example the difference between the C$\rightarrow$T and T$\rightarrow$C rates, with Bayes factors.
The fact that a model with the C$\rightarrow$T and T$\rightarrow$C rates equal (reversible wrt C$\leftrightarrow$T) is nested within the random-effects model allows us to use the Savage-Dickey ratio \citep[\textit{e.g.}][]{wagenmakers2010bayesian} to compute the Bayes factor from the posterior distribution of the random-effects model, as we discuss in the Supplemental Section ``Assessing the strength of evidence for nonreversibilities.''
(There is no need to fit any additional models or estimate marginal likelihoods directly.)
The Bayes factor provides ``very strong'' \citep{kass1995bayes} support for the nonreversibility of C$\rightarrow$T and G$\rightarrow$T rates (over the reversible model).
We can also assess the strength of evidence for nonreversibilities via the posterior sign probability.
This is the posterior probability that the sign of a variable is the same as the sign of the posterior median \citep[this is one minus the tail probability used by][]{zhang2021large}.
The estimated sign probability ranges from 0.5 to 1.0, with larger values indicating increasingly strong support that the parameter is non-zero.
Here, as with Bayes factors, we are interested in the sign probabilities of the differences in random-effects rather than the random-effects directly.
The sign probabilities agree with the Bayes factors that there is strong evidence for the nonreversibility of C$\rightarrow$T and G$\rightarrow$T rates, with both estimated sign probabilities above 0.99.

\begin{figure}[h]
	\centering
	\includegraphics[width=0.75\textwidth]{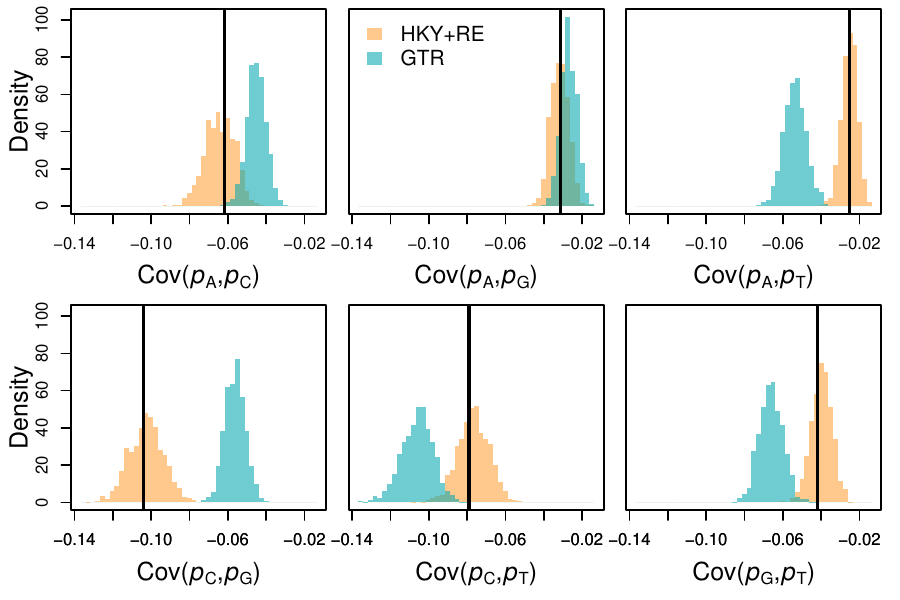}
	\caption{
	Posterior predictive distributions of the covariances of the proportions of each nucleotide (denoted $p_A$, $p_C$, $p_G$ and $p_T$) across sites in the alignment (histograms) compared to the true values (vertical black lines).
  	The HKY+RE predictive distributions all closely align with the observed values while all but one of the GTR predictive distributions are discordant.
	}%
	\label{fig:sc2_pps}
\end{figure}

Given the strong evidence for nonreversibilities, we sought to investigate the issue of the adequacy of reversible models (namely GTR) using posterior predictive model checks.
In a posterior predictive framework, a summary of the observed dataset is compared to the distribution of summaries of datasets produced by drawing from the posterior distribution on model parameters.
Broadly, if the model fits the data well, we expect that the predicted summaries will match the observed values, while if the fit is poor there will be a mismatch.
As our test statistics, we consider all pairwise covariances of the proportion of each nucleotide (A, C, G, and T) across the alignment (we discuss this in more detail in the Supplemental Section ``Posterior predictive p-values for proportions'').
These test statistics clearly demonstrate that the HKY+RE model better captures the evolutionary processes at hand (Figure~\ref{fig:sc2_pps}).
Compared to inference using GTR, the analysis with HKY+RE produces notably higher support for the root-most divergence (the 95\% credible set includes 67 possible resolutions for GTR and 1 for HKY+RE, Supplemental Figure S8) and infers a root time approximately 5 days earlier.

\subsection{Phylogeography of influenza from 2002-2007}

For a larger state-space example of random-effects substitution models, we consider the global spread of human influenza A virus (subtype H3N2) from 2002 to 2007.
\citet{lemey2014unifying} examined the movement patterns between 14 distinct air travel communities using 1529 viral genomes.
The authors used a generalized linear model (GLM) to parameterize the spread of the virus between these communities as a function of a number of covariates, and discovered that the most consistently supported predictor of spread between communities was the volume of air traffic.

We re-analyze this dataset using a GLM substitution model with random-effects.
We now briefly review the setup of a GLM substitution model, and our random-effects extension.
For each pair of locations $i$ and $j$, let $\bX_{ij}$ be a vector of $P$ predictors of the rate of movement from $i$ to $j$ (these may depend on the source $i$, the destination $j$, or both) with associated coefficients $\bbeta$.
A GLM substitution model with random-effects defines the rate matrix through
\begin{equation}
	\log \ratematrixelement_{ij}  = \bX_{ij}' \bbeta + \epsilon_{ij}.
\end{equation}
This is a log-linear model, in which the GLM defines a substitution rate based on predictors and the estimated coefficients, and the random-effects allow for deviations from the model's predictions.

In particular, we employ a simple GLM with only air traffic included as a predictor.
This approach allows us to determine how well air traffic volume predicts the spread of influenza A virus in the mid-2000s.
If most random-effects are negligible, then air traffic volume alone is perhaps adequate for modeling the spread of influenza in this time frame.
On the other hand, if many or most random-effects are not negligible, although air traffic volume may be an important model component, it is not sufficient to explain spread, absent random-effects.
While \citet{lemey2014unifying} used spike-and-slab priors on $\bbeta$ in a Bayesian model averaging approach, since we are using only predictors identified previously to be important, we use a Normal prior instead (corresponding to the slab in the original study).
We apply Bayesian bridge priors for the random-effects.
To account for phylogenetic uncertainty, we marginalize our inference over the same empirical distribution phylogenetic trees used by \citet{lemey2014unifying}.

\begin{figure}[ht]
	\centering
	\includegraphics[width=0.67\textwidth]{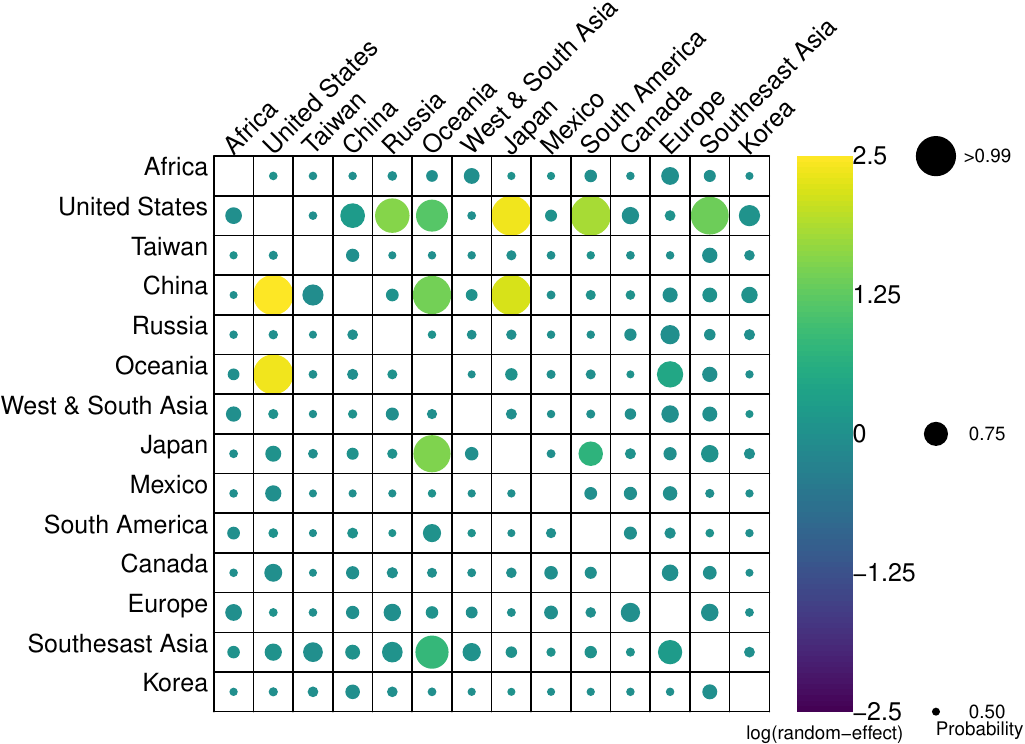}
	\caption{
	Summary of all 182 random-effects for the influenza A virus (subtype H3N2), shown in the format of the rate matrix, with the source in rows and destination in columns.
	The circle in each square is colored by the posterior median random-effect.
	The size of the circle denotes how strong the posterior support is that a random-effect is in the model.
	Specifically, the radius corresponds to the posterior sign probability.
	When the prior dominates the posterior distribution, a random-effect gains a larger posterior mass at 0 and becomes increasingly symmetric, the median approaches 0, and the posterior sign probability approaches 0.5.
	When the data are strongly informative, the posterior distribution moves away from 0 and the posterior sign probability gets larger.
	The random-effects which are most strongly supported are all positive, indicating that air travel underpredicts dispersal for those pairs of locations.
	}%
	\label{fig:flu_refx}
\end{figure}

We find that air traffic volume sufficiently explains the viral spread between most communities.
That is, for most community pairs, the posterior distribution of the random-effect indicates that the parameter has been declared ``insignificant'', and is a spike centered at 0 (Figure~\ref{fig:flu_refx}).
However, for 5 pairs of communities (from the United States to Japan and South America; from China to the United States and Japan; from Oceania to the United States), the inferred random-effect is clearly significant (all sign probabilities $>0.99$) and strongly positive, indicating 6- to 12-fold higher dispersal than predicted by travel.
There is support for an additional 6 random-effects (from the United States to Oceania, Russia, and Southeast Asia; from China to Oceania; from Japan to Oceania; from Southeast Asia to Oceania) which have sign probabilities between 0.87 and 0.97 and correspond to 2- to 5-fold higher dispersal than predicted by travel.
All other area pairs of sign probabilities are less than 0.78.

Given the offset seasons between hemispheres, some of these connections likely do not represent biologically meaningful connections, and may potentially be attributed to sampling biases.
A comparison of the number of samples in the dataset to the population sizes of the regions (a rough proxy for the number of infections in the regions) reveals that the United States, Oceania, and Japan are strongly oversampled.
Thus, sampling biases likely explains many of the significant random-effects, including the between-hemisphere connections.
As China is not particularly oversampled, the elevated rates of transmission from China may represent source-sink dynamics which are not captured by air travel alone, rather than sampling bias.

\subsection{Analysis of paired macroevolutionary traits}

Random-effects substitution models can also be used to test for dependent substitution processes between multiple characters as follows.
Let us assume we have two characters of interest, $\bY_1$ and $\bY_2$.
These characters could be morphological, behavioral, or even ecological traits.
If these characters evolved independently along the phylogeny $\phylogeny$, we could model this with two rate matrices, $\ratematrix^1$ and $\ratematrix^2$, a (strict) clock rate which defines the rate of change (in substitutions per year or million years) for $\bY_2$, and a relative rate parameter $\mu$ which defines how much faster (or slower) $\bY_1$ evolves compared to $\bY_2$.
We can define a composite character from $\bY_1$ and $\bY_2$ by considering both states simultaneously.
This yields a new character $\bY$ which is the cartesian product of the two state-spaces, with the combined state-space size $\alphabetsize = \alphabetsize_1 \times \alphabetsize_2$.
The rate matrix $\ratematrix$ for the combined character is 0 for any double substitution and for any single substitution is defined by $\ratematrix^1$ or $\ratematrix^2$ depending on which character changes.
Written on the log-scale, the (unnormalized) rate matrix is given by
\begin{equation}
	\log \ratematrixelement_{ij}  = 
	\begin{cases}
    \log \ratematrixelement^1_{ij}   & \mbox{if i$\rightarrow$j substitution in $Y_1$}\\
		\log \mu + \log \ratematrixelement^2_{ij}  & \mbox{if i$\rightarrow$j substitution in $Y_2$}\\
		-\infty & \mbox{for all double substitutions}.\\
	\end{cases} \nonumber
\end{equation}
We can test for departures from independent evolution by allowing the state of one character to modulate the rates of change between states in the other through the addition of random-effects.

We employ this random-effects dependent morphological evolution model on a dataset of 29 species of frogs in the family Hylidae (subfamily Hylinae).
Taking the phylogeny inferred by \citet{caviedes2020species} to be fixed, we focus on two traits described in \citet{caviedes2019}, one ecological and one behavioral.
The ecological trait is the habitat, which is characterized as either arboreal or understory.
The behavioral trait is the swimming mode, which is characterized by whether the back legs move in an alternating or simultaneous fashion or whether both types are observed.
To determine the structure of the underlying independent-trait models, we first fit the independent model using asymmetric rates for both traits.
Bayes factors show no evidence for any model more complex than the Mk (Jukes-Cantor-like) model \citep{lewis2001likelihood}.

In particular, we are interested in whether the degree of arboreality, defined as habitat preference, impacts the swimming mode, as canopy-dwelling species move the back legs in an alternating fashion while climbing.
Thus, we place random-effects only in the direction of arboreality affecting swimming mode.
Letting $\bY_1$ be arboreality and $\bY_2$ be swimming mode, the unnormalized rate matrix for our random-effects substitution model is,
\begin{equation}
	\log \ratematrixelement_{ij} = 
	\begin{cases}
    0 & \mbox{if i$\rightarrow$j substitution in $Y_1$}\\
    \log \mu + \randomeffect_{ij} & \mbox{if i$\rightarrow$j substitution in $Y_2$ and $Y_1$ indicates canopy-dwelling}\\
	\log \mu & \mbox{if i$\rightarrow$j substitution in $Y_2$ and $Y_1$ indicates understory-dwelling}\\
	-\infty & \mbox{for all double substitutions}.\\
	\end{cases} \nonumber
\end{equation}

\begin{figure}[h]
	\centering
	\includegraphics[width=0.75\textwidth]{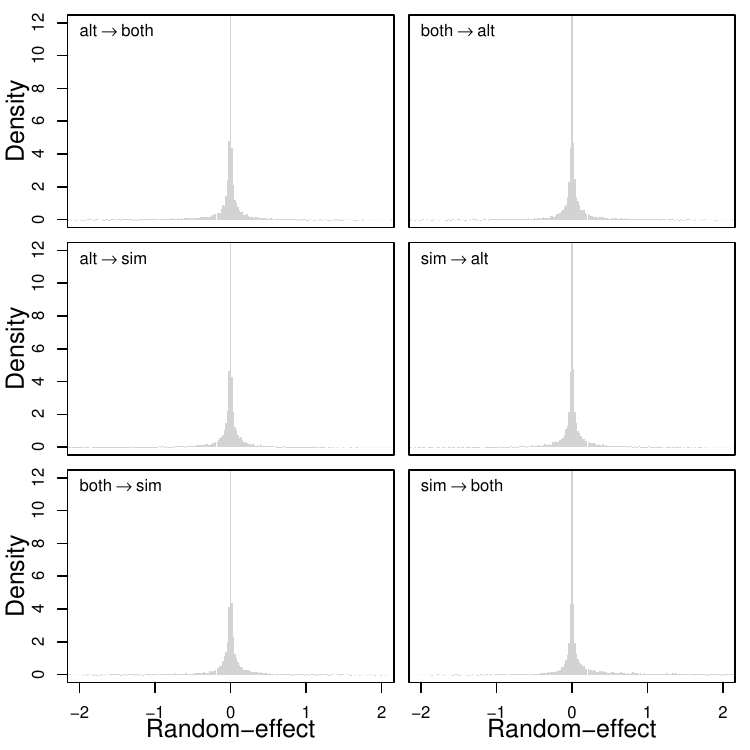}
	\caption{
	Posterior distributions of the six random-effects which capture the effect of arboreality on swimming mode.
	Abbreviations are ``alt'' for back legs moving in an alternating fashion, ``sim'' for back legs moving in a simultaneous fashion, and ``both'' for both types of movements.
	Each row groups forward and reverse transitions.
	The shape of the posterior distributions is strongly indicative of parameters which have been shrunk out of the model via the Bayesian bridge prior, and all posterior sign probabilities are less than 0.57.
	}%
	\label{fig:frog_posteriors}
\end{figure}

We infer no effect of arboreal habitat on the swimming mode, all posterior sign probabilities are between 0.5 and 0.57, indicating that all random-effects have clearly been deemed insignificant (Figure~\ref{fig:frog_posteriors}).
We also infer that the rate of habitat evolution is roughly twice that of swimming-mode evolution ($\mu \approx 0.5$, 95\% CI 0.23-1.16).
There are two important caveats to these results.
First, with only 29 species, the power to detect dependent evolution is likely low unless the effect is quite large.
Secondly, by only modeling two traits, we are missing out on possible interactions between other aspects of ecology (such as the aquatic environments the species make use of) and morphology (such as the lengths of limbs and digits) which might modulate this relationship.

\subsection{Performance gains from gradients}

\begin{figure}[!htp]
	\centering
	\includegraphics[width=0.7\textwidth]{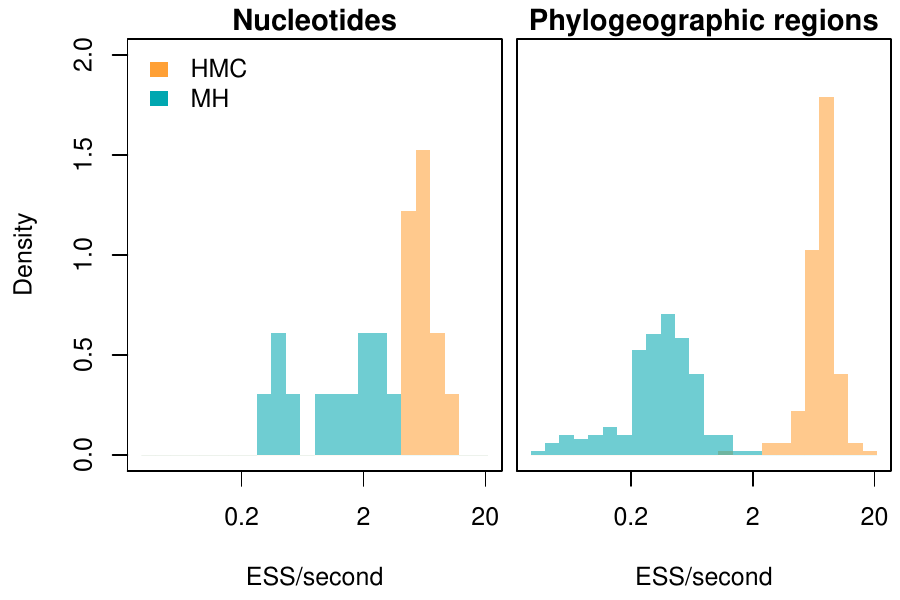}
	\caption{
		Efficiency, in effective samples per second (ESS/s), of HMC versus MH-MCMC for inferring the random-effects substitution models for both the 12 random-effects in an HKY+RE model and the 182 random-effects in the flu discrete phylogeographic analysis.
		Both data sets show markedly improve estimation efficiency as a result of employing HMC with approximate gradients.
		}%
	\label{fig:ess_per_second}
\end{figure}

For inferring random-effects in nucleotide substitution models, we find a notable improvement in efficiency using HMC with our approximate gradients over using standard random-walk MH-MCMC.
For our measure of efficiency, we consider the number of effectively independent samples taken per second (ESS/s).
This measure incorporates both the increased ability of HMC to move through parameter space, as well as the increased cost per MCMC move required for repeated evaluation of the gradient.
We track the efficiency separately for each random-effect (that is, we use the univariate ESS), and we consider two summaries of efficiency gains from HMC.
As an overall measure of efficiency increase, we consider the parameterwise average increase in the efficiency.
However, as analyses are constrained by waiting for the slowest-mixing parameter to achieve a sufficiently large ESS, we also consider the improvement in the minimum ESS (regardless of which parameter is slowest-mixing).
When applied to nucleotide models (HKY+RE) to infer the tree from sequence data, we observe an average parameterwise increase in efficiency of 6.6 fold, and an increase in the minimum efficiency of 14.8-fold (Figure~\ref{fig:ess_per_second}).
For the larger state-space of the flu phylogeographic example (14 discrete areas), where we average across a set of posterior samples of the tree from the original study, we find an average parameterwise increase in efficiency of 20.2-fold and an increase in minimum efficiency of 33.6-fold (Figure~\ref{fig:ess_per_second}).
Timing was done on a Macbook Pro with an 8-core CPU M1 Pro chip and 32GB of memory.

\subsection{Analyses of simulated data}
To assess the performance of random-effects substitution models in estimation of model parameters, we performed a simulation study.
We based the simulation setup on our analysis of the SARS-CoV-2 data.
In particular, we used the posterior distribution of trees, the HKY $\kappa$ parameter, and the shape parameter governing the Gamma-distributed among-site rate variation.
For the random-effects, we simulated from three groups: null effects, moderate effects, and strong effects.
For each of these we drew values from Normal distributions (Supplemental Figure S3) chosen to reflect the values observed in the real-data posterior distributions.
The strong effects were C$\rightarrow$T and G$\rightarrow$T, which were simulated from a Normal(1.50,0.18) distribution.
The moderate effects were A$\rightarrow$T and G$\rightarrow$A, which were simulated from Normal(-0.68,0.37) and Normal(0.68,0.37) distributions respectively.
The remaining 8 random-effects were classified as null and simulated from Normal(0.0,0.11) distributions.
We simulated 100 datasets under this model.

Analyses of simulated datasets were conducted following the analysis of the SARS-CoV-2 dataset, with two exceptions.
First, we treated the tree as known.
Second, we considered several values for the exponent parameter, each simulated dataset was analyzed four times with $\bridgeExponent = 1/8,1/4,1/2,1$.
These values range from strongly-regularizing priors (small $\bridgeExponent$) to the weakly-regularizing Laplace prior ($\bridgeExponent = 1$).

\begin{figure}
	\centering
	\includegraphics[width=0.6\textwidth]{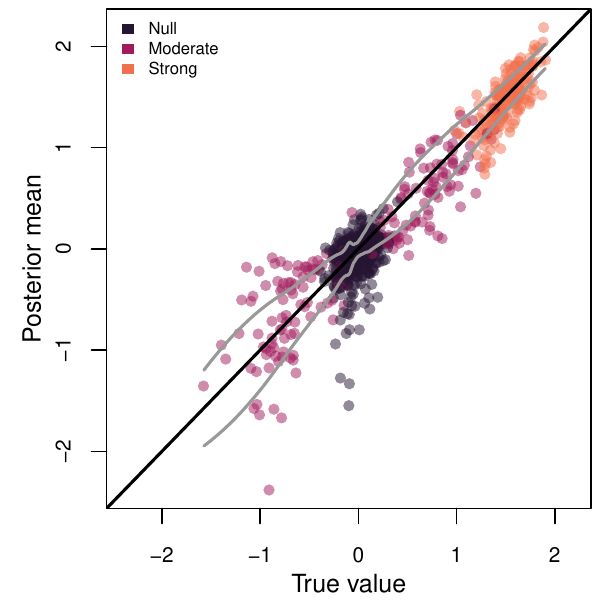}
	\caption{
	Estimation performance of HKY+RE on simulated data.
	Posterior mean parameter value versus true value simulated, colored by whether the true value was drawn from the distribution on null, moderate, or strong effects.
	As most posterior means are near the true value (close to the solid black line), parameter estimation is generally good.
	The solid grey lines display locally-smoothed estimates of the standard deviation of the error $\hat{\theta} - \theta_{\text{true}}$, showing that null and strong effects are better-estimated than moderate ones.
	}%
	\label{fig:simulation_study_results}
\end{figure}

Overall, we find that random-effects are well-estimated and that random-effects which imply strong deviations from the base model (HKY) can be confidently identified using the posterior sign probability.
Overall the posterior means are strongly correlated with the true simulating value, ($\rho = 0.95$), though it appears that null and strong effects are generally better-estimated than moderate effects (Figure~\ref{fig:simulation_study_results} and Supplemental Figure S5).
To determine whether a random-effect is significant, one can use a threshold on the sign probability, declaring larger sign probabilities to be evidence for significant effects.
Particularly notable deviations from the base model are easy to detect at any chosen threshold.
Lower thresholds declare many negligible deviations to be significant, while higher thresholds are somewhat underpowered to detect smaller, but potentially important, deviations.
A threshold of around 0.8 (0.75 to 0.85) provides a good balance between these forces (Supplemental Figure S4).

Examining the choice of exponent $\bridgeExponent$, we find that values on the order of $1/4$ provide a reasonable trade-off between estimation performance and MCMC behavior.
The smaller exponents, $1/8$ and $1/4$, in general produce posteriors which are notably closer to the true values than the larger values (Supplemental Figure S5).
The difference in performance is less notable for the evidence for significance provided by sign probabilities.
The smaller exponents perform better for identifying null effects as null, while the larger exponents produce more confident support that moderate effects belong in the model, and all coefficients do well with strong effects (Supplemental Figure S6).
We investigated MCMC efficiency by examining the minimum effective sample size per sample.
This efficiency is higher at $\bridgeExponent = 1/4$ than any other exponent (Supplemental Figure S7).

\section{Discussion}
In this paper, we demonstrated the versatility and usefulness of random-effects substitution models.
By wrapping around a simpler base substitution model, random-effects substitution models enable increased flexibility while retaining the useful structure of the base model.
Applied to a dataset of 583 SARS-CoV-2 sequences, an HKY+RE model picks up strong C$\rightarrow$T and G$\rightarrow$T mutational biases and is shown by posterior predictive model checks to be an adequate substitution model where reversible models like GTR fail.
Used with a GLM substitution model to analyze the phylogeographic pattern of spread of influenza in humans, the random-effects suggest the air traffic volume alone is a powerful explanation for the spread of influenza from 2002 to 2007.
In examination of the evolution of ecological and behavioral characters in hylid tree frogs, a random-effects model shows no evidence for an effect of arboreality on the mode of swimming.
Simulations show that random-effects can be accurately estimated and provide guidelines for interpreting whether a random-effect is significant or not.

To enable efficient inference of random-effects substitution models, we derived an approximate substitution gradient.
The time-complexity of our approximate approach is cubic in the size of the state-space, while ``exact'' analytical techniques are quintic.
For parameter-rich random-effects substitution models, numerical gradients are also quintic, and our approximate gradients enable \textit{maximum a posteriori} inference of the parameters of an amino-acid substitution model over 50 times faster than numerical gradients (Section ``Inferring the dynamics of amino acid substitution in Metazoa'' \onDryad).
Used in Bayesian inference, we find that HMC using our approximate gradients is 6.6 to 20.2 times more efficient than standard Metropolis-Hastings moves, with yet more substantial gains when comparing the dimension with the most difficult sampling (where the efficiency gains are 14.8 and 33.6 fold).
In particular, it appears that the efficiency of HMC with the approximate gradients is roughly invariant to the dimension (Figure~\ref{fig:ess_per_second}).
For our SARS-CoV-2 example, with a $4 \times 4$ rate matrix, the average efficiency of HMC is 6.3 effective samples per second, while for the influenza A virus phylogeographic example, with a $14 \times 14$ rate matrix, it is 7.6 effective samples per second.
However, the efficiency of Metropolis-Hastings moves decreases from 1.5 effective samplers per second to 0.41.
We expect this trend to continue as the size of the state-space increases, and that for sufficiently large models (such as codon models or Markov-modulated amino acid models), HMC will be the only approach capable of inferring random-effects substitution models in any reasonable timeframe.

Although the approximate substitution gradient we derived performed very well in our applications, it cannot be expected to perform ideally in every circumstance.
Mathematical analysis and in-silico experiments suggest that the error in our approximation grows with the branch length measured in genetic distance (Section ``Error in the approximate gradient'' \onDryad).
Thus, we should expect performance to be best where the tree has few substitutions per site.
\citet{wertheim2022accuracy} refer to this as the near-perfect regime, and it is common in viral phylodynamic applications.
However, we note reasons for optimism in applying our approximate gradients in regimes with larger numbers of substitutions.
The influenza phylogeographic example falls outside the near-perfect regime, and the efficiency of HMC using our approximate gradients is still quite good.
Similarly good inference efficiency is observed in \textit{maximum a posteriori} inference of an amino acid model on a Metazoan tree which has over 5.5 substitutions per site on average (Section ``Inferring the dynamics of amino acid substitution in Metazoa'' \onDryad).
It is also important to note that when used for HMC, the accept-reject step ensures correctness even in regimes where the approximation gets poor.
It is likely the error bound we have obtained is quite conservative.
Further, \citet{didier2023surprising} establish a more rigorous error bound and show that the error decreases with increasingly large state-spaces, suggesting that phylogeographic analyses are well-suited to this approximation.

An open question is to define the regimes where the approximation becomes poor enough that inference becomes inefficient such that other techniques would be preferable.
We note two such alternative approaches which could be considered and compared to the efficiency of our approximation in future work. 
An exact gradient can be obtained from a data-augmentation procedure which jointly samples the complete mutational history along the tree, such as the approach adopted by \citet{lartillot2006conjugate}.
Within the framework of approximate gradients, an affine correction, as \citet{didier2023surprising} suggest, may yield smaller approximation error when the expected number of substitutions per branch is relatively large.

There are a number of important extensions of this work.
Currently, we have implemented the gradient computations (in BEAST 1.10 \citep{suchard2018bayesian} and BEAGLE 3 \citep{ayres2019beagle}) for use on CPUs, however GPU-based likelihood computations have proven incredibly efficient in many phylogenetic contexts \citep{suchard2009many,dudas2017virus,ayres2019beagle,baele2021markov,lemey2021untangling}.
In particular, \citet{gangavarapu2023many} recently showed minimum increases of 8-fold and 128-fold for nucleotide and codon models respectively when computing gradients with respect to branch rate parameters.
A GPU implementation of our approximate gradients would likely produce notable speedups in inference, especially for large state-space models.
Mathematically, our approximation holds for any case in which there is a single substitution rate matrix on any edge of the phylogeny (though we have currently only implemented the case for a single rate matrix across the whole tree).
However, the process of geographic spread may be temporally inhomogeneous while applying consistently across all lineages alive at any given time.
In such cases, epoch models \citep{bielejec2014inferring,gao2022new} are needed.
The epoch times break branches into multiple regimes, which requires matrix convolutions for likelihood computation and thus an extension of our approach.

Random-effects substitution models are a flexible approach for creating more realistic substitution models, but they are not a panacea.
They cannot, for example, address gross violations of the underlying assumptions of the CTMC model, such as memorylessness.
Nor can they address dependence between characters without carefully predefining the set of (potentially) coevolving characters and expanding the state space of the model.
The Bayesian bridge provides a robust framework for regularization, and HMC an efficient framework for inference.
However, the additional complexity of random-effects models may occasionally cause challenges for MCMC which require more active user intervention.
Consider, for example, the (likely) APOBEC-induced C$\rightarrow$T bias observed in our SARS-CoV-2 example, which, in a double-stranded virus, will also lead to a G$\rightarrow$A bias \citep{gigante2022multiple,otoole2023putative}.
Application of HKY+RE to such a dataset will lead to multimodality (caused by ridges in the likelihood) jointly involving five substitution model parameters, $\kappa$ and the pairs of random-effects $\randomeffect_{G \rightarrow A},\randomeffect_{C \rightarrow T}$ and $\randomeffect_{A \rightarrow G},\randomeffect_{T \rightarrow C}$.
Such multimodality does not invalidate the model, and it could be mitigated by the use of TN93+RE or avoided entirely by using a simpler model like F81+RE.

\section{Data and code availability}
BEAST XML files for the analyses in this paper are available at, as well as the supplementary text, are available on Dryad, \url{https://doi.org/10.5068/D1709N}.
The approximate gradients have been implemented in the hmc-clock branch of BEAST (\url{https://github.com/beast-dev/beast-mcmc/tree/hmc-clock/}) and the v4.0.0 release of BEAGLE (\url{https://github.com/beagle-dev/beagle-lib/releases/tag/v4.0.0}).
BEAST XML files for the analyses in this paper are additionally available at \url{https://github.com/suchard-group/approximate_substitution_gradient_supplement}.

\section{Acknowledgments}
This work was supported through NSF grants DMS 2152774 and DMS 2236854, as well as NIH grants R01 AI153044, R01 AI162611 and K25 AI153816.
J. O. W. was funded by AI135992.
J. E. P. was funded by NIH T15LM011271.
Dr.\ Matsen is an Investigator of the Howard Hughes Medical Institute.
P. L. acknowledges funding from the European Research Council under the European Union's Horizon 2020 research and innovation programme (grant agreement no. 725422-ReservoirDOCS) and from the European Union's Horizon 2020 project MOOD (grant agreement no. 874850).
G. B. acknowledges funding from the Internal Funds KU Leuven under grant agreement C14/18/094, from the Research Foundation -- Flanders (`Fonds voor Wetenschappelijk Onderzoek -- Vlaanderen', G0E1420N and G098321N) and from the DURABLE EU4Health project 02/2023-01/2027, which is co-funded by the European Union (call EU4H-2021-PJ4) under Grant Agreement No. 101102733.
We gratefully acknowledge support from Advanced Micro Devices, Inc. with the donation of parallel computing resources used for this research.

\bibliographystyle{plainnat}
\bibliography{ms}

\clearpage

\beginsupplement

\section{Inferring the dynamics of amino acid substitution in Metazoa}\label{sec:metazoa}
To test our approximate gradients in an optimization setup, we require an alternative approach for computing gradients.
In particular, we use numerical gradients which represent a gold standard for assuring the correctness of gradient implementations.
For random-effects substitution models the computational complexity of numerical gradients, $\bigO(\nTaxa \alphabetsize^5)$, is the same as the non-approximate gradient with mapping presented in Equation 14, making it an appropriate comparison for speed as well.
For our example, we seek to estimate the parameters of an amino acid substitution model.
We use a 28 taxon, 445 site amino alignment (locus OG198) from \citet{borowiec2015extracting} which spans the Metazoan tree of life.
We fix the authors' inferred maximum-likelihood tree for the locus \citep{borowiec2015dryad}, and following the authors' choice of best-fit amino acid model, we fit an LG \citep{le2008improved} substitution model with random-effects.

\begin{figure}
	\centering
	\includegraphics[width=0.5\textwidth]{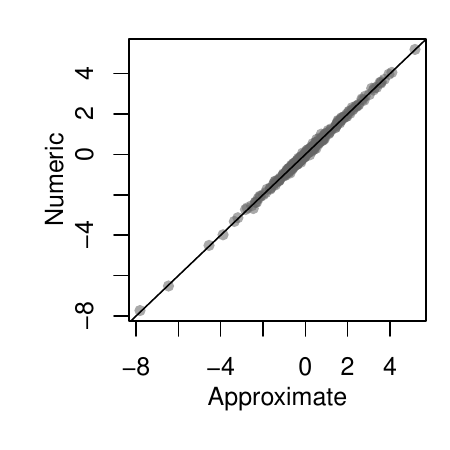}
	\caption{
		Comparison of MAP estimates of random effects in a LG+RE model when estimated using approximate and numeric gradients.
	}%
	\label{fig:optimization}
\end{figure}

We perform MAP estimation of the model parameters.
We compare approximate MAP inference using the approximate likelihood gradient derived in this paper to correct MAP inference using numerical (finite-difference) likelihood gradients.
Using numerical gradients, the L-BFGS optimizer requires 1.19 seconds per iteration, while using the approximate gradients 0.023 seconds are required per iteration.
Thus, the approximate gradients are computed over 50 times faster for estimates that are essentially indistinguishable (Figure~\ref{fig:optimization}).
(Timing for optimization was also done on a Macbook Pro with an 8-core CPU M1 Pro chip and 32GB of memory.)
It is also possible to combine both the approximate and numerical gradients.
That is, first we could run until convergence with the approximate gradient, which is faster than the numerical gradient.
Then, we could use the more accurate numerical gradient to get the exact maximum.

Once the posterior mode has been found, it is possible to employ a Laplace approximation \citep{kass1995bayes} and approximate the posterior distribution as a multivariate normal around the mode.
The covariance matrix is estimated as the inverse of the (negative) Hessian matrix.
This allows us to approximate the posterior probability that a coefficient is non-zero.
However, as the numerical gradients do not produce an invertible Hessian, the Laplace approximation for this analysis is potentially unstable and we do not report any results for it.

\section{More details on models employed}
For completeness, we now present a more details of the analyses in the paper, including the supplemental analysis of Metazoan amino acid evolution, in Table~\ref{tab:data}.

\begin{table}[H] 
    \centering
    \begin{tabular}{l | c c c c} \toprule
		\multicolumn{1}{r|}{\textbf{Dataset}} & SARS-CoV-2 & Influenza A virus & Hylidae & Metazoa \\
	 \midrule
	 \multicolumn{1}{c|}{\textit{Tree}} &  &  & & \\
	 \textbf{Number of taxa} & 583 & 1529 & 29 & 28 \\
	 \textbf{Inference} & Inferred & Sampled$^1$ & Fixed & Fixed \\
	 \textbf{Time-calibrated} & Yes & Yes & Yes & No \\
	 \midrule
	 \multicolumn{1}{c|}{\textit{Random-effects}} &  &  &  &  \\
	 \textbf{Data type} & DNA & Geography & Mixed$^2$ & Protein \\
	 \textbf{Number} & 12 & 182 & 6 & 380 \\
	 \midrule
	 \multicolumn{1}{c|}{\textit{Base model}} &  &  & & \\
	 \textbf{Model} & HKY & GLM$^3$ & Mk-like$^4$ & LG \\
	 \textbf{Base frequencies} & Empirical & Equal & Equal & Fixed$^5$ \\
	 \textbf{Parameters inferred} & $\kappa$ & One fixed-effect$^3$ & Relative rate $\mu$$^4$ & -- \\
	 \midrule
	 \multicolumn{1}{c|}{\textit{Alignment}} &  &  & & \\
	 \textbf{Data type} & DNA & -- & -- & Protein \\
	 \textbf{Alignment length} & 29,903 & -- & -- & 445 \\
	 \textbf{Site patterns} & 1034 & -- & -- & 393 \\
	\bottomrule
    \end{tabular}
    \caption{%
	Details of the empirical datasets used in this study, inapplicable values are denoted by --.
	$^1$The influenza A virus tree is sampled from the empirical distribution of \citet{lemey2014unifying}.
	$^2$There is one behavioral character and one ecological character.
	$^3$The phylogeographic GLM for influenza A virus dispersal includes only a single fixed-effect representing the impact of air travel, as measured by the number of seats on scheduled commercial flights which is the same as in \citet{lemey2014unifying}.
	$^4$The base model for the Hylid ecological and behavioral evolution is equivalent to independent Mk models on both characters separately with one parameter specifying the relative (strict) clock rate difference between them.
	$^5$The base frequencies are fixed to the LG default values.
	}
    \label{tab:data}
\end{table}

\section{More on Hamiltonian Monte Carlo with surrogate trajectories}

Recent independent works \citep{neklyudov2020involutive,glatt2020accept,andrieu2020general} have provided an alternative justification (to that presented in the main text) for the use of approximate gradients in HMC.
To wit, these works provide mathematical foundations for the use of \emph{involutions}, or deterministic mappings $\mathbf{i}:\mathbb{R}^{2K}\rightarrow \mathbb{R}^{2K}$ that satisfy $\mathbf{i}\left(\mathbf{i}(\params,\mom)\right)=(\params,\mom)$, within general MCMC algorithms.
This framework can greatly simplify the determination of whether a potential MCMC algorithm targets the correct distribution.

Briefly, following \citet{neklyudov2020involutive}, involutive MCMC algorithms work in an expanded state space including both the parameters $\params$ and an auxiliary variable $\mom$ which have a joint density $p(\params,\mom)$.
A new state is proposed first by drawing a value of the auxiliary variable $\mom$ and then using the involution to obtain the proposed values of both values, $(\params^*,\mom^*) = \mathbf{i}(\params,\mom)$.
This state is accepted with probability $\max(1,p(\params^*,\mom^*)/p(\params,\mom) \times \left| \nabla \mathbf{i}(\params,\mom) \right|)$, where $\left| \nabla \mathbf{i}(\params,\mom) \right|$ is the determinant of the Jacobian of the involution.
(For an alternative overview under the name ``Metropolis-Hastings-Green algorithm with Jacobians'', see \citet{vats2023hamiltonian}.)
For HMC, the auxiliary variable is the momentum and the targeted density is defined by the exponential of the Hamiltonian energy function ($H(\params,\mom) = \log(p(\params,\mom))$).
So for HMC, the above three steps are: first, draw the initial momentum; second, obtain the proposed state by simulating from the Hamiltonian dynamics given that initial momentum; third, accept or reject the proposed state.

Such theoretical results relate to the present work insofar as the combination of leapfrog dynamics (Equation 19) with a sign-flip $(\params,\mom) \mapsto (\params,-\mom)$ constitutes an involution.
To see this, suppose one has performed a single leapfrog iteration followed by a sign-flip.
Next, starting at time $\tau+\epsilon$, a subsequent leapfrog iteration unwinds the first:
    \begin{align}\label{eq:leap2}
	\mom(\tau + \frac{3}{2}\epsilon) &= -\mom(\tau+\epsilon) + \frac{\epsilon}{2} \nabla \log \pi(\params(\tau+\epsilon))  \\  \nonumber
	\params(\tau+2\epsilon) &= \params(\tau+\epsilon) + \epsilon \, \cov^{-1}\mom(\tau + \frac{3}{2}\epsilon) \\ \nonumber
	\mom(\tau + 2\epsilon) &= \mom(\tau + \frac{3}{2}\epsilon) + \frac{\epsilon}{2} \nabla \log \pi(\params(\tau+2\epsilon))  \,.
\end{align}
One may verify that $(\params,\mom)(\tau+2\epsilon)=(\params,\mom)(\tau)$ by substituting individual terms, e.g.,
\begin{align*}
	\params(\tau+2\epsilon) &= \params(\tau+\epsilon) + \epsilon \, \cov^{-1}\mom(\tau + \frac{3}{2}\epsilon) \\
	&= \left(\params(\tau) + \epsilon \, \cov^{-1}\mom(\tau+\epsilon/2) \right) + \epsilon \, \cov^{-1}\left( -\mom(\tau+\epsilon) + \frac{\epsilon}{2} \nabla \log \pi(\params(\tau+\epsilon))  \right) \\
	&=\left(\params(\tau) + \epsilon \, \cov^{-1}\mom(\tau+\epsilon/2) \right) +\\
	&\quad \epsilon \, \cov^{-1}\left( -\left(\mom(\tau+\epsilon/2) + \frac{\epsilon}{2} \nabla \log \pi(\params(\tau+\epsilon))  \right) + \frac{\epsilon}{2} \nabla \log \pi(\params(\tau+\epsilon))  \right) \\
	&= \params(\tau) \, .
\end{align*}
One may similarly show that $\mom(\tau+2\epsilon)=\mom(\tau)$, and analogous results immediately follow for the composition of $L$ leapfrog steps with a momentum sign-flip.
Importantly, this algebra remains the same when one substitutes an arbitrary function $\mathbf{g}$ for the log-posterior gradient (say, an approximation using $\approxnabla \likelihoodAll{\params}$ instead of $\nabla \likelihoodAll{\params}$), and the upshot is a deeper theoretical justification for the use of surrogate gradients within HMC.

\section{Computational complexity of alternative approaches to computing the gradient of a matrix exponential}\label{sec:supplemental_computational_complexity}

In this paper, we have employed a first-order approximation to the derivative of the matrix exponential with respect to its elements, $\frac{\partial}{\partial \ratematrixelement_{ij}} \exp(t \ratematrix)$.
In this section, we contemplate the computational efficiency of alternative means of evaluating relevant gradients more exactly.
First, we consider a more exact approach to evaluating the gradient of a matrix exponential with respect to its elements.
Then, we consider finite differences to directly attack the gradient of the phylogenetic log-likelihood.

\subsection{A less approximate approach}
Following \cite{najfeld1995derivatives}, let us define the block-matrix Z,
\begin{align*}
  \bZ =
  \begin{bmatrix}
    \ratematrix & \indicatormatrix_{ij}\\
    0 & \ratematrix
  \end{bmatrix}
\end{align*}
Then, we have (Equation 90, \cite{najfeld1995derivatives}),
\begin{equation}
  \exp(t \bZ) =
  \begin{bmatrix}
    \exp(t \ratematrix) & \frac{\partial}{\partial \ratematrixelement_{ij}} \exp(t \ratematrix)\\
    0 & \exp(t \ratematrix)
  \end{bmatrix}
  \label{eqn:nh_ex\postPartial_algorithm}
\end{equation}
Thus on a single branch $\node$ for the $\bigO(\alphabetsize^3)$ cost of exponentiating one $2 \alphabetsize \times 2 \alphabetsize$ matrix, we obtain both the matrix exponential and its derivative with respect to a single element of the rate matrix, $\partial \exp(t \ratematrix) / \partial \ratematrixelement_{ij}$.
However, we will need the full matrix of all such partial derivatives, $\allQderivs_{\node}$, meaning we need to repeat the process for all $\alphabetsize^2$ elements of $\ratematrix$, which is $\bigO(\alphabetsize^5)$.
Summing the per-branch contributions across the tree and employing the mapping procedure of the Section ``Reducing the computational complexity'' allows us to obtain the gradient in $\bigO(K \alphabetsize^2 + \nTaxa \alphabetsize^5)$.
This cost is prohibitively large for even moderately-sized $\alphabetsize$.

\subsection{The numerical approach}\label{sec:numerical_approach}
The gradient of the log-likelihood can also be obtained numerically via finite differences.
To approximate $\dBydParam \likelihoodAll{\params}$ using finite differences, we simply need to evaluate the likelihood twice.
If $\indicatorvector_k$ is a vector in which all elements are 0 except the $k$th, to obtain the central finite difference we need to evaluate $\likelihoodAll{\params + (h/2) \indicatorvector_k}$ and $\likelihoodAll{\params - (h/2) \indicatorvector_k}$, where $h$ is a very small quantity.
The numerical approximation is then
\[
  \dBydParam \likelihoodAll{\params} = \frac{\big( \likelihoodAll{\params + (h/2) \indicatorvector_k} - \likelihoodAll{\params - (h/2) \indicatorvector_k} \big)}{h}.
\]
This is the numerical approach employed in the Section ``Performance gains for MAP estimation.''

To evaluate the log-likelihood, a matrix exponential must be evaluated on each of $\bigO(\nTaxa)$ branches on the tree.
When computing the exponential of an $\alphabetsize \times \alphabetsize$ matrix by eigendecomposition, the requirements of diagonalization and matrix multiplication make the operation $\bigO(\alphabetsize^3)$ \citep{suchard2009many}.
This makes the numerical gradient for a single parameter $\bigO(\nTaxa \alphabetsize^3)$, and the gradient $\bigO(K \nTaxa \alphabetsize^3)$.
Recall that the cost of our approximate gradients using parameter mappings is $\bigO(K \alphabetsize^2 + \nTaxa \alphabetsize^3)$.
If the number of parameters $K$ in the rate matrix is small, then numerical approaches may be viable.
Among commonly-used GTR-family models, gradient-based inference for the K2P model would likely be more efficient with numerical derivatives than the approaches outlined in this paper.
However, any model with unequal base frequencies (\textit{e.g.} moving from K2P to HKY) has $K \gtrsim \alphabetsize$, such that the numerical approach is expected to be less efficient than the approximate approach with parameter mappings.
For a richly-parameterized random-effects model where every element of $\ratematrix$ has an effect and there are roughly $\alphabetsize^2$ parameters, the approximate gradient with mapping is clearly more efficient.
As the state-space increases in size, the magnitude of the increase in efficiency of the approximate gradient will become larger.
This is corroborated by the 50-fold speed increase of MAP optimization using the approximate gradient over numeric gradients reported in the Section ``Performance gains for MAP estimation.''

\section{Error in the approximate gradient}\label{sec:error}

As we are approximating the gradient, one natural question is, how good is the approximation?
We start by splitting Equation 15 into two parts, the portion of the sum $\bS$ which we use for the first-order approximation, and the remainder term $\bR$
\begin{align}
	\label{eqn:error_norm}
	\frac{\partial}{ \partial \ratematrixelement_{ij}} e^{\ratematrix \branchlength} &= e^{\ratematrix \branchlength} \sum_{x=0}^{\infty} \frac{\branchlength^{x+1}}{(x+1)!} \{\indicatormatrixij,\ratematrix^x\}\\ \nonumber
	&= e^{\ratematrix \branchlength} \Big( \branchlength \indicatormatrixij + \sum_{x=1}^{\infty} \frac{\branchlength^{x+1}}{(x+1)!} \{\indicatormatrixij,\ratematrix^x\} \Big) \\ \nonumber
	&= e^{\ratematrix \branchlength} \big( \bS + \bR \big). \nonumber
\end{align}
Revisiting Equations 13 and 17 with this formulation in mind, we have that the true gradient is
\begin{align}
	\allQderivselement_{\node i j} &= \bpostPartial_{\node \category \site}' \frac{\partial}{ \partial \ratematrixelement_{ij}} \bprePartial_{\node \category \site} \nonumber \\
	&= \bpostPartial_{\node \category \site}' \left( \frac{\partial}{\partial \ratematrixelement_{ij}} e^{\ratematrix \times t_{\node} \times \rateCategory_{\category}} \right)' \bprePrePartial_{\node \category \site} \nonumber \\
	&=  \bpostPartial_{\node \category \site}' \left(  e^{\ratematrix \times t_{\node} \times \rateCategory_{\category}} \left( \bS + \bR \right) \right)' \bprePrePartial_{\node \category \site}  \nonumber \\
	&=\bprePartial_{\node \category \site}' (\bS + \bR) \bpostPartial_{\node \category \site},
\end{align}
while our approximation is
\begin{align}
	\allQderivselement_{\node i j} &\approx \bprePartial_{\node \category \site}' \, \bS \, \bpostPartial_{\node \category \site}.
\end{align}  

To understand how far our approximation deviates from the truth, we must quantify the remainder term $\bR$.
Using matrix norms, we will examine its magnitude, $\mnorm \bR \mnorm$ and the perhaps more-informative relative magnitude $\mnorm \bR \mnorm / \mnorm \bS \mnorm)$.
We will require two matrix norm identities.
First, we have (from repeated application of the fact that matrix norms obey the triangle inequality) that,
\[
 \bigmnorm\sum_i \bA_i\bigmnorm \leq \sum_i \mnorm\bA_i\mnorm.
\]
Second, we have for matrix commutator series $\{\bA,\bB^n\}$ that,
\[
 \mnorm\{\bA,\bB^n\}\mnorm \leq 2^n \times \mnorm A \mnorm \times \mnorm B \mnorm,
\]
which can be found in \citet{najfeld1995derivatives}.
From here on, for simplicity of notation, we drop per-branch subscripts and use $t$ for the product $t_{\node} \times \rateCategory_{\category}$.

Now, starting with our definitions of $\bS$ and $\bR$ from Equation~\ref{eqn:error_norm} and our first identity, we have
\begin{align*}
	\mnorm \bR \mnorm &=  \Biggmnorm \sum_{x=0}^{\infty} \frac{\branchlength^{x+1}}{(x+1)!} \{\indicatormatrixij,\ratematrix^x\} \Biggmnorm \\ \nonumber\\
  &\leq \sum_{x=1}^{\infty} \Biggmnorm \frac{\branchlength^{x+1}}{(x+1)!} \{\indicatormatrixij,\ratematrix^x\} \Biggmnorm \\ \nonumber
	&\leq  \sum_{x=1}^{\infty} \frac{\branchlength \times \mnorm \indicatormatrixij \mnorm \times \mnorm \bQ \mnorm}{x+1} \frac{(2 \branchlength)^{x}}{x!} \\ \nonumber
	&= (\branchlength \times \mnorm \indicatormatrixij \mnorm \times \mnorm \bQ \mnorm) \sum_{x=1}^{\infty} \frac{1}{x+1} \frac{(2 \branchlength)^{x}}{x!}\\
  &= \frac{\mnorm \indicatormatrixij \mnorm \times \mnorm \bQ \mnorm}{2}(e^{2t} - 2t - 1).
\end{align*}
We can also note that,
\[
  \mnorm \bS \mnorm = t \mnorm \indicatormatrixij \mnorm,
\]
meaning that we can bound the relative magnitude $\mnorm \bR \mnorm / \mnorm \bS \mnorm)$ as,
\[
  \frac{\mnorm \bR \mnorm}{\mnorm \bS \mnorm} \leq \frac{\mnorm \bQ \mnorm}{2t}(e^{2t} - 2t - 1)
\]

Note that $\branchlength$ and $\mnorm \bQ \mnorm$ are strictly non-negative.
The bounds on both the absolute and relative error depend on the norm of the rate matrix and get larger as the branch length (measured in genetic distance) increases.
Thus we should expect approximate gradient should perform best when branches are short, as in the near-perfect regime of \citet{wertheim2022accuracy}.
We note also that this is not a particularly tight error bound.

\subsection{Numerical experiments}
\begin{figure}[htp]
	\centering
	\includegraphics[width=0.9\textwidth]{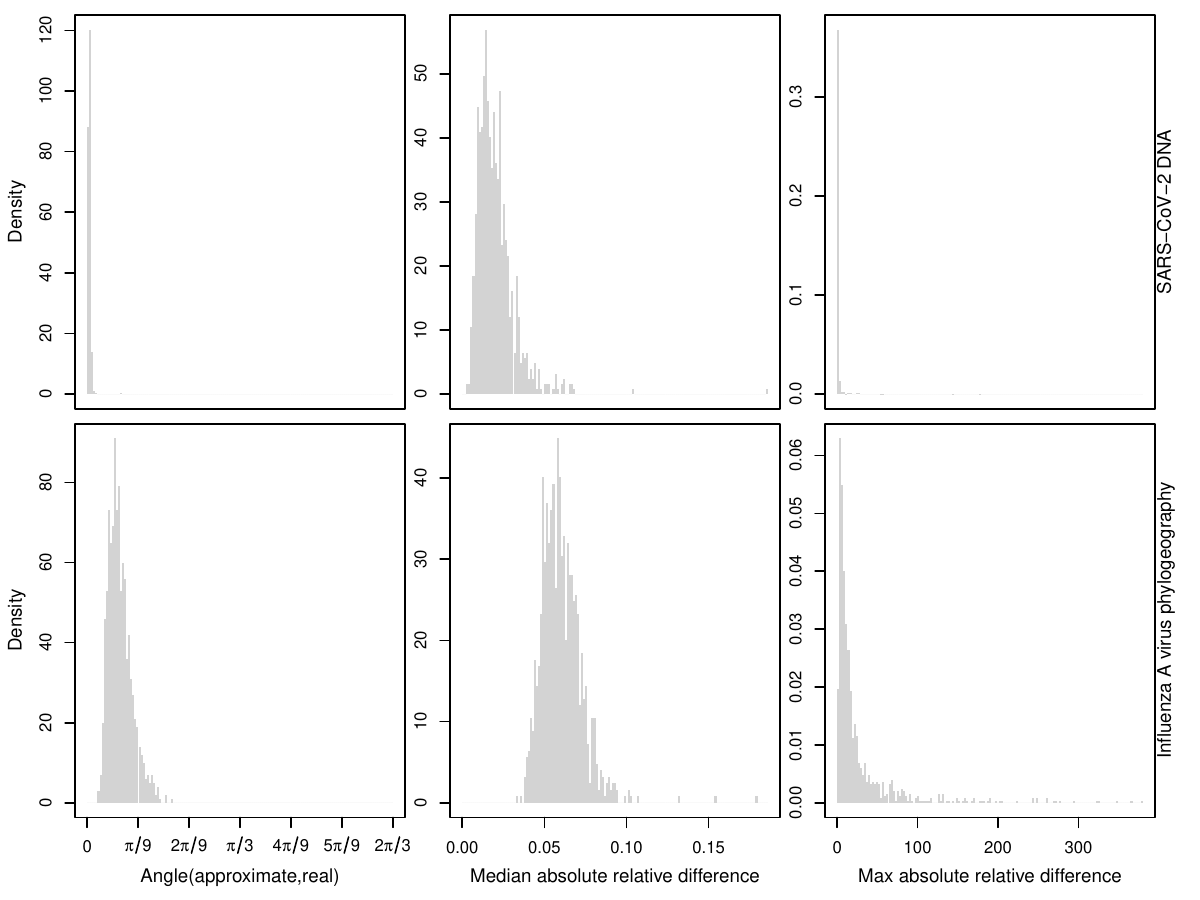}
	\caption{
		Error in the approximate gradients accross a set of 1000 samples from the joint posterior of the SARS-CoV-2 molecular evolution example (top) and influenza A virus phylogeographic example (bottom).
		Different columns use different summaries of the error in each posterior sample: the angle between the approximate and true gradient vectors, the elementwise median absolute difference (relative to the true gradient), and the elementwise maximum absolute difference (relative to the true gradient).
		For the rightmost column, the largest 1\% of values have been excluded for interpretability.
		In all cases, numerical approaches are used to obtain the true gradient.
	}%
	\label{fig:gradient_error}
\end{figure}

A less general, but potentially more enlightening, approach for quantifying the error in our approximate gradients is to test it experimentally.
In particular, we can compare our approximate gradient to numerical gradients across a set of posterior samples.
The numerical gradients represent a gold standard stand-in for the true gradients, while restricting ourselves to the (joint) posterior distribution on model parameters tells us what the error looks like in regions of parameter space which have non-negligible support.
As expected, we see that the error is worse for the influenza A phylogeographic analysis, where the tree is longer in terms of the total number of substitutions per site, than the SARS-CoV-2 analysis (Figure~\ref{fig:gradient_error}).
However, while the maximum elementwise difference grows quite large in the phylogeographic example, the median shows that most parameters' partial derivatives are much closer, and the angle between the approximate and true gradient vectors stays relatively small.

\section{Assessing the strength of evidence for nonreversibilities}\label{sec:savage}
In some cases, such as examining the geographic spread of influenza A virus, it will be of interest to assess the strength of evidence that a particular random-effect is non-zero.
In other cases, such as our SARS-CoV-2 example, however, the question is instead about the evidence for asymmetry in the substitution model.
In this case, we are instead interested in the distribution of $\randomeffect_{ij} - \randomeffect_{ji}$ and the evidence that this is non-zero.
To assess the evidence for nonreversibilities, non-zero random-effects, or other similar questions (such as strand-symmetry), we can use Bayes factors.
In particular, as our models are nested, we may use the Savage-Dickey ratio to obtain the Bayes factor.

In the case where we are interested in asymmetry, we must first re-parameterize our model.
Previously, we wrote the log-scale random-effects extension of a base model $\basemodel$ with instantaneous rate matrix $\boldsymbol{B} = \{ b_{ij} \}$, as,
\[
  \log(\ratematrixelement_{ij}) = \log(b_{ij}) + \randomeffect_{ij}.
\]
Instead we will write the log-scale rate matrix elements for reverse substitution directions $i \rightarrow j$ and $j \rightarrow i$ together, as,
\begin{align}
  \ln(\ratematrixelement_{ij}) &= \log(b_{ij}) + \randomeffect_{ij} \nonumber \\
  \ln(\ratematrixelement_{ji}) &= \log(b_{ji}) + \randomeffect_{ij} + \Delta_{ij}.  \nonumber
\end{align}
If $\basemodel$ is reversible, then its random-effects extension ($\basemodel$+RE) will be reversible when $\Delta_{ij} = 0$.
In this case, we will not have changed the ratio between the rate matrix elements, $\ratematrixelement_{ij}/\ratematrixelement_{ji} = b_{ij}/b_{ji} = \pi_j / \pi_i$.

The distribution on $\Delta_{ij}$ is the distribution of the difference of two Bayesian-bridge-distributed variables (namely $\randomeffect_{ji} - \randomeffect_{ij}$), such that we can define $\randomeffect_{ji} = \randomeffect_{ij} + \Delta_{ij}$ and recover the model as written previously.
The Bayes factor in favor of $\Delta_{ij} = 0$ (Model 0, against Model 1 where $\Delta_{ij}$ is a free parameter) is the ratio of the posterior density to the prior density at $\Delta_{ij} = 0$,
\[
  \text{BF}_{01} = \frac{\pr(\Delta_{ij} = 0 \mid \by)}{\pr(\Delta_{ij} = 0)}
\]

The posterior distribution will not in general have a known closed-form solution, so kernel density estimation will be needed to estimate $\pr(\Delta_{ij} = 0 \mid \by)$ from posterior samples.
The prior density $\pr(\Delta_{ij} = 0)$ can be written in closed form if we are not employing shrunken-shoulder regularization on the Bayesian bridge.
In the case where shrunken-shoulder regularization is employed and the tails of the distribution are much lighter, we can run our model under the prior and use kernel density estimation to estimate the prior density at 0.

Let us now derive the (non-regularized) Bayesian bridge density at 0.
Denote the Bayesian bridge density with global scale $\globalScale$, exponent $\bridgeExponent$, and no shrunken-shoulder regularization as $f_{\text{BB}}(x;\globalScale,\bridgeExponent)$.
Let $Z(\globalScale,\bridgeExponent)$ be the normalizing constant for the Bayesian bridge with those parameters.
The probability density of $\Delta_{ij}$ at 0 is given by,
\begin{align*}
  f_{\Delta_{ij}}(0) &= \int_{-\infty}^{\infty} f_{\text{BB}}(x;\globalScale,\bridgeExponent) f_{\text{BB}}(-x;\globalScale,\bridgeExponent) dx \\
  &= \frac{1}{Z(\globalScale,\bridgeExponent)^2} \int_{-\infty}^{\infty} \exp({-|x/\globalScale|^\bridgeExponent}) \exp({-|-x/\globalScale|^\bridgeExponent}) dx\\
  &= \frac{1}{Z(\globalScale,\bridgeExponent)^2} \int_{-\infty}^{\infty} \exp({-2|x/\globalScale|^\bridgeExponent}) dx.
\end{align*}
We can recognize the integrand as the kernel of a Bayesian bridge distribution with global scale $2^{-1/\bridgeExponent} \globalScale$ and exponent $\bridgeExponent$.
Thus we have,
\[
  f_{\Delta_{ij}}(0) = \frac{Z(2^{-1/\bridgeExponent} \globalScale,\bridgeExponent)}{Z(\globalScale,\bridgeExponent)^2}.
\]
The Bayesian bridge distribution (without shrunken-shoulder regularization) is also known as the Exponential Power Distribution and the Generalized Normal Distribution and has normalizing constant \citep{griffin2018gnorm},
\[
  Z(\globalScale,\bridgeExponent) = \frac{\bridgeExponent}{2 \globalScale \Gamma(1/\bridgeExponent)}.
\]
Thus, the prior density at $\Delta_{ij} = 0$ is,
\[
  f_{\Delta_{ij}}(0) = \frac{2^{1 + 1/\bridgeExponent} \globalScale \Gamma(1/\bridgeExponent)}{\bridgeExponent}
\]
When the global scale is a parameter in the model (as it is in all of our applications), this can be numerically integrated over the prior on $\globalScale$.
In the case where shrunken-shoulder regularization is employed, the marginal Bayesian bridge distribution is instead proportional to $\exp({-|x/\globalScale|^\bridgeExponent}) \exp(-x^2/(2\xi^2))$ and the normalizing constant is not known.

\section{Posterior predictive p-values for proportions}\label{sec:PPS}
For completeness, we now write out explicitly how to compute the posterior predictive summaries used to compare HKY+RE to GTR on the SARS-CoV-2 data in the Section ``C to T bias in SARS-CoV-2 evolution.''
Recall that our summaries treat the columns as observations of the proportion of nucleotides at each site.
We use proportions, rather than counts, such that the statistic can be computed comparably across all sites for all non-ambiguous nucleotide states.
That is, ambiguities can be ignored, as long as the same sites are masked out in the posterior predictive alignment.

Denote the alignment as $\bY$, it has $\nTaxa$ rows and $\nSites$ sites, and $\bY_j$ is a single column.
Let $\mathcal{A}$ be the alphabet (of size $\alphabetsize$) and $\mathcal{A}_i$ be a particular character in the alphabet.
We define a new $\nSites \times \alphabetsize$ matrix, $\bp$, of per-site proportions of characters by,
\begin{equation}
  \postPartial_{ij} = \frac{\sum_{k=1}^{\nTaxa} \indicator(y_{ki} = \mathcal{A}_j)}{\sum_{l=1}^{\alphabetsize} \sum_{k=1}^{\nTaxa} \indicator(y_{ki} = \mathcal{A}_l)}.
\end{equation}
That is, the rows in this new matrix are the sites in the alignment, and the columns are the proportions (among all non-ambiguous characters) of each character at that site.

We now have $\alphabetsize$ new variables, $\bp_1,\dots,\bp_{\alphabetsize}$, stored as columnns in $\bp$.
We restrict our attention to variable sites, such that the number of rows in $\bp$ is the number of variable sites.
For DNA the alphabet consists of the nucleotides A, C, G, and T, and $\bpostPartial$ is a $\nSites \times 4$ matrix.
The summaries of interest are the means, variances, and covariances of these new variables.
The means, give or take discrepancies from ignoring ambiguities, are the proportions of each of the characters in the alignment.
Thus, they should largely reflect a model's ability to capture large-scale features of the substitution process, like the stationary frequencies of a GTR model.
The variances should, at least partially, reflect the tree length.
For a tree of length 0, each site is exclusively one character, and each site is essentially a draw from a categorical distribution with probabilities given by the root frequency distribution.
For a tree where all branches are of infinite length, each site is a draw from an $\nTaxa$-dimensional multinomial distribution with probabilities given by the equilibrium frequencies.
The (finite, non-zero) length of the tree will determine where along this continuum our variances fall.
The covariances describe the strength of association of two characters.
We might expect if a model misses an extremely large rate, such as $C \rightarrow T$, it might under-estimate the corresponding covariance.
Though since normalization (and potentially assumptions of symmetry) bind the rates together, the effect may cascade and lead to over-estimation of other covariances.

\section{Maximization-based approaches}
While this paper is primarily interested in Bayesian inference via HMC, the approximate gradients we have derived are also useful in maximization-based approaches to estimation such as maximum likelihood (ML) estimation and maximum \textit{a posteriori} (MAP) estimation.
Let us call the function we wish to optimize $f(\params)$ that is a function of our model parameters $\params$.
If $f(\params) = - \log \likelihoodAll{\params}$, then minimizing $f(\params)$ is equivalent to maximizing the likelihood and our maximum-likelihood parameter estimates are
$
  \hat{\params} = {\argmin}_{\params}\ f(\params)
$.
MAP estimation instead uses $f(\params) = - \log \likelihoodAll{\params} - \log \pr(\params)$ and maximizes the joint posterior density.

\subsection{Optimization routines in BEAST}\label{sec:bfgs}
Given access to $f(\params)$ and its gradient $\nabla f(\params)$, a variety of algorithms exist for numerically finding the minimum.
BEAST 1.10 \citep{suchard2018bayesian} offers users access to the L-BFGS optimization algorithm, a limited-memory version of the BFGS algorithm \citep[see, \textit{e.g.}][]{dennis1996numerical,ji2020gradients}.
Both the BFGS and L-BFGS algorithms use information about the curvature of the likelihood surface from the Hessian (the matrix of second derivatives) to guide the search,
At each step in the algorithm $t$, the L-BFGS algorithm uses the current gradient $\nabla f(\params_{t})$, and an approximation to the inverse Hessian $\boldsymbol{H}_{t}$ to define a direction of descent $\boldsymbol{p}_t = \boldsymbol{H}_{t} \nabla f(\params_{t})$.
Then a line-search is used to choose the step size $\alpha_{t}$ that minimizes $f(\params_{t} + \alpha_{t} \boldsymbol{p}_t)$, and the parameters are updated accordingly, $\params_{t+1} = \params_{t} + \alpha_{t} \boldsymbol{p}_t$.
These steps are repeated until the minimum is obtained.
Where the BFGS algorithm stores and updates the entire approximate inverse Hessian $\boldsymbol{H}_{t}$, the L-BFGS algorithm stores only a recent history of iterations.
At each step, this history is used to implicitly carry out operations requiring $\boldsymbol{H}_{t}$, reducing the computational complexity by reducing the requisite number of matrix multiplications.

By using our approximate gradient $\approxnabla \log \likelihoodAll{\params}$ in place of the true gradient $\nabla \log \likelihoodAll{\params}$, impressive speed gains may be realized.
Optimization solely using  $\approxnabla \log \likelihoodAll{\params}$ produces approximate inference; however, a final round of optimization based on $\nabla \log \likelihoodAll{\params}$ may be performed to often obtain exact estimates without requiring extensive use of the more expensive true gradient.

\section{Simulation study}\label{sec:simulations}
We performed a simulation study to quantify the performance of random-effects substitution models in estimation of model parameters and to examine the impact of the choice of the exponent parameter $\bridgeExponent$.
Our simulation setup is based heavily on our analysis of the SARS-CoV-2 data, using the posterior distribution for the phylogeny, the HKY $\kappa$ parameter, and the shape parameter governing the Gamma-distributed among-site rate variation.

\begin{figure}[htp]
	\centering
	\includegraphics[width=0.5\textwidth]{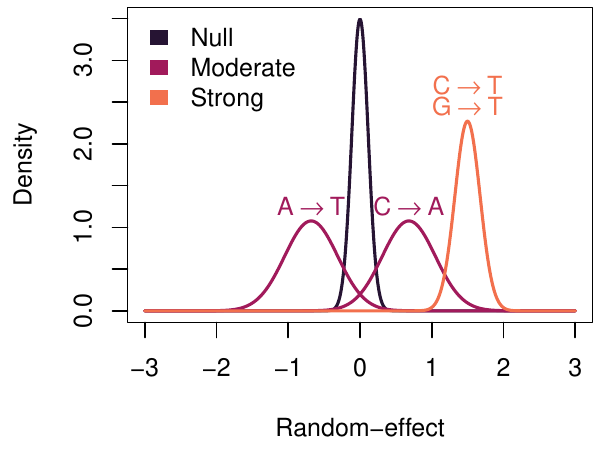}
	\caption{
		Normal distributions used for simulating the random-effects.
	}%
	\label{fig:normals}
\end{figure}

In order to make the results more interpretable, we did not simulate directly from the posterior distribution on the random-effects.
Instead, we fit Normal distributions to three classes of posterior distributions, those which were clearly shrunk to 0 (null effects), those which were estimated unambiguously to be in the model (strong effects, A$\rightarrow$T and G$\rightarrow$A), and those in between (moderate effects, C$\rightarrow$T and G$\rightarrow$T).
These are shown in Figure~\ref{fig:normals}.

We simulated 100 datasets under our Normal-distributed random-effects model.
We analyzed each of these using four different choices of the exponent parameter $\bridgeExponent \in (1/8,1/4,1/2,1)$.
This led to 400 analyses of simulated data.
As this is too many analyses to manually inspect for convergence, we used a fully-automated procedure.
Burnin was determined by considering windows of 10\%, 20\%, 30\%, 40\%, and 50\% of the chain and picking the burnin that produced the highest ESS of the trace of the joint posterior density.
This procedure trades off between choosing too small of a burnin (which reduces the ESS by increasing autocorrelations at large time lags) and too large a burnin (which reduces the ESS by reducing the sample size).
Any run with an effective sample size below 200 was discarded entirely.

\begin{figure}[htp]
	\centering
	\includegraphics[width=0.75\textwidth]{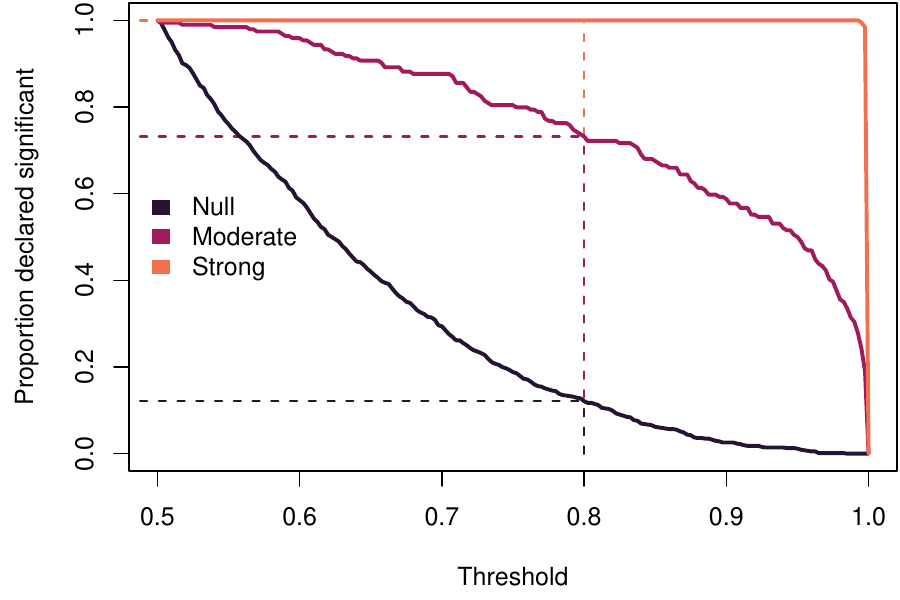}
	\caption{
		The proportion of random-effects declared significant from simulated data analyses as a function of the significance threshold.
		While all strong effects are rightly declared significant at essentially any chosen threshold, there is a clear trade-off between the ability to detect moderate effects and the (incorrect) declaration of null effects to be significant.
		At a threshold of 0.8, 100\% of strong effects, 73\% of moderate effects, and 12\% of null effects are declared significant (dashed lines).
	}%
	\label{fig:sign_prob_threshold_choice}
\end{figure}

When choosing a sign probability threshold to use for declaring random-effects significant, we must strike a balance between true and false positives.
We want a threshold which keeps the proportion of negligible effects declared significant acceptably small, while allowing the proportion of non-negligible effects detected to be sufficiently large.
In Figure~\ref{fig:sign_prob_threshold_choice}, we examine the proportion of null, moderate, and strong effects detected as a function of the threshold, for our chosen exponent parameter $\bridgeExponent = 1/4$.

In the main text, we presented estimation performance results for both the average Euclidean distance between the posterior distribution and the true effect and the posterior sign probability for our chosen exponent parameter of $\bridgeExponent = 1/4$.
For completeness, we now present simplified figures showing the performance for all parameter values considered.
We also present results for MCMC performance.

\begin{figure}[htp]
	\centering
	\includegraphics[width=0.75\textwidth]{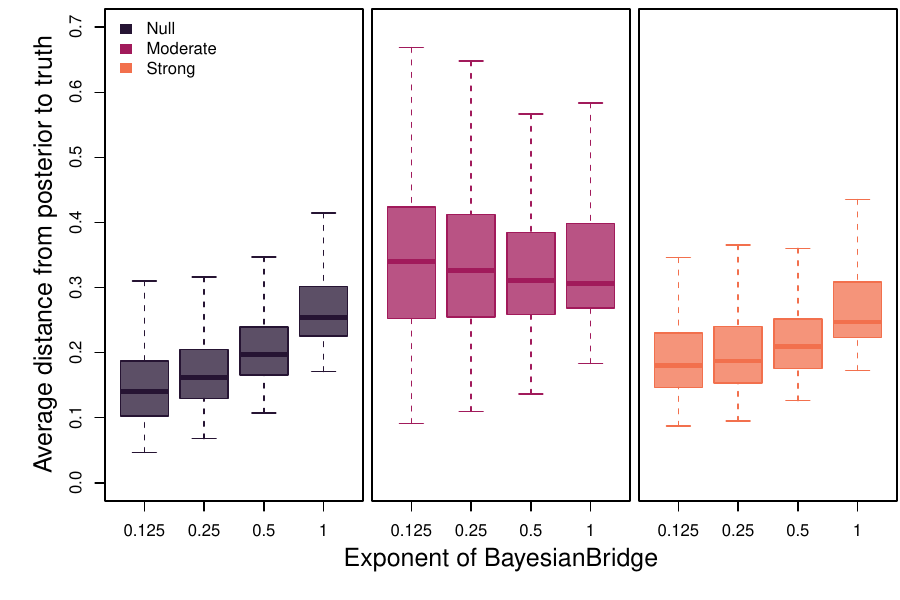}
	\caption{
		Average distance from the posterior distribution to the true coefficient on simulated datasets.
		Lower values denote posteriors which are on average closer to the truth.
		For both null and strong effects, performance is better at smaller values of the exponent parameter $\bridgeExponent$, while for moderate values performance does not vary as strongly.
		Colors of coefficients show null (black, left panel), moderate (purple, center panel), and strong (orange, right panel) coefficients.
	}%
	\label{fig:estimation_performance_supp}
\end{figure}

\begin{figure}[htp]
	\centering
	\includegraphics[width=0.75\textwidth]{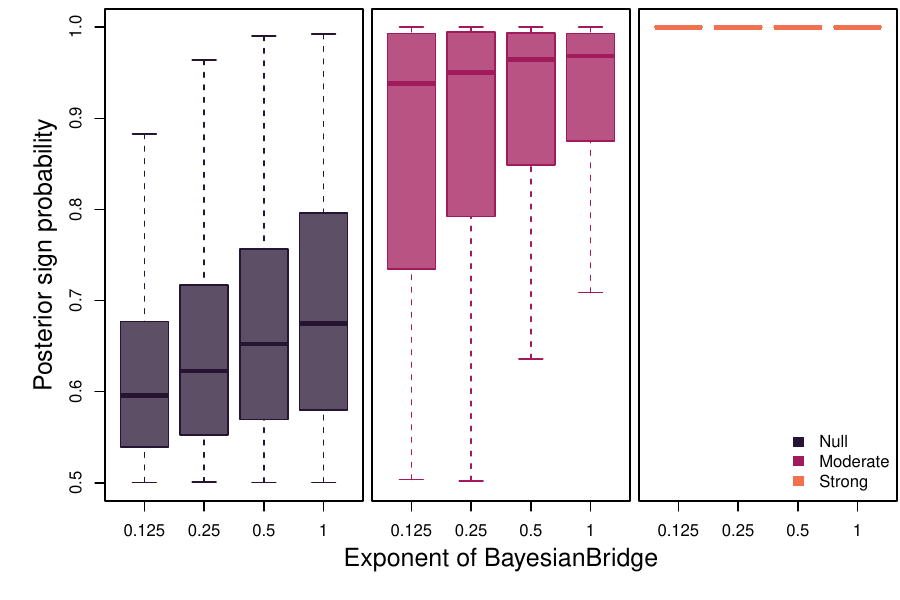}
	\caption{
		Posterior sign probability for simulated datasets.
		Values near 0.5 indicate maximum uncertainty about the sign, corresponding to clearly null random-effects, while values close to 1.0 indicate maximum certainty, corresponding to clear evidence the random-effect belongs in the model.
		For null coefficients, smaller values of the exponent parameter $\bridgeExponent$ lead to better classification of null coefficients as null.
		For moderate coefficients, performance is better with larger coefficients.
		Colors of coefficients show null (black, left panel), moderate (purple, center panel), and strong (orange, right panel) coefficients.
	}%
	\label{fig:sign_probs_supp}
\end{figure}

\begin{figure}[htp]
	\centering
	\includegraphics[width=0.75\textwidth]{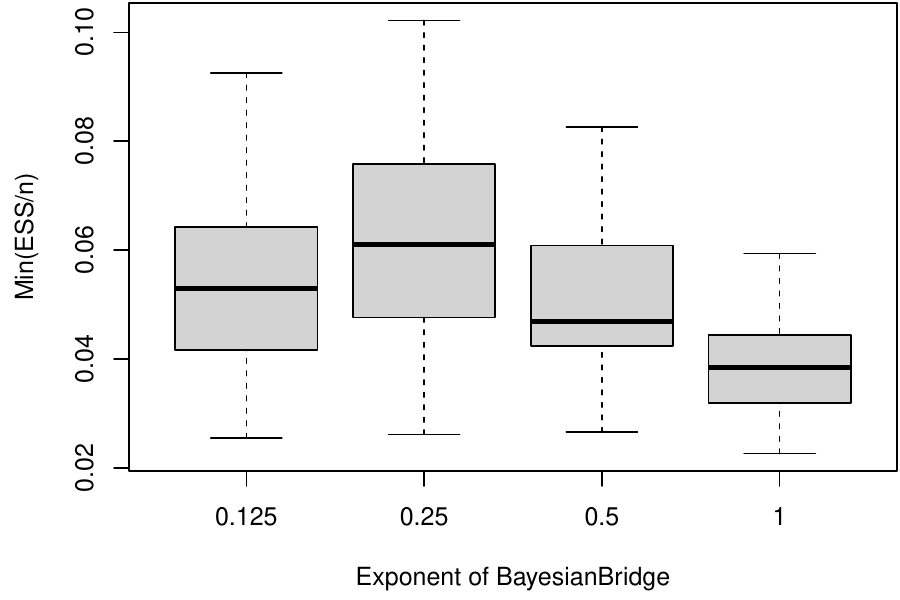}
	\caption{
		MCMC efficiency as measured by the minimum (across all values in the log file)of the effective sample size per sample (ESS/n).
		The parameter with the lowest ESS defines how long a run must be in order to have an acceptably low Monte Carlo standard error.
		Efficiency is best at $\bridgeExponent = 1/4$.
	}%
	\label{fig:exponent_mixing}
\end{figure}

\clearpage
\section{Resolution at the root of the SARS-CoV-2 tree}
\begin{figure}[htp]
	\centering
	\includegraphics[width=0.75\textwidth]{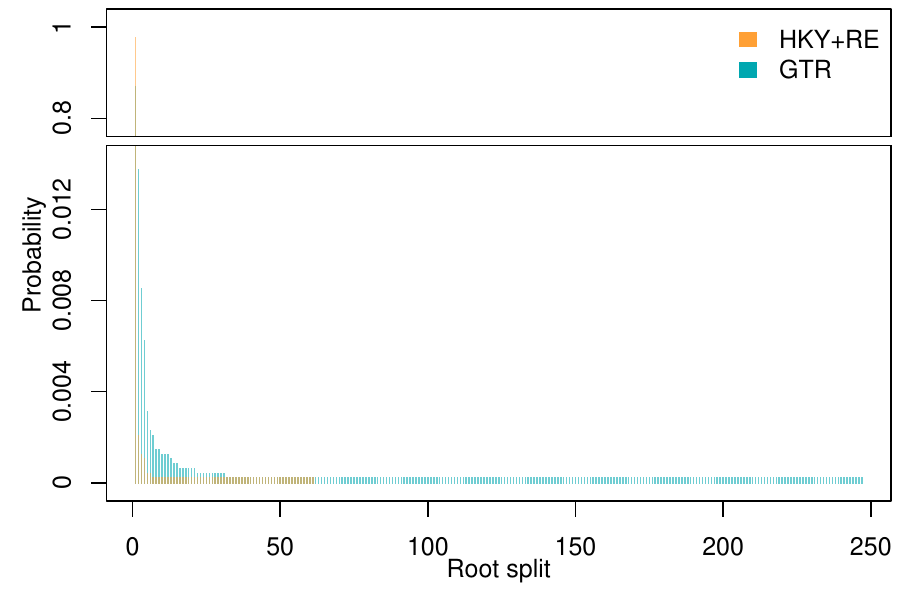}
	\caption{
		Support for different root splits for the SARS-CoV-2 dataset when analyzed using GTR and HKY+RE.
		Analysis with HKY+RE provides higher posterior probability to the most-probable root split than GTR and samples far fewer splits in total.
		The axis is split to provide better resolution for the remaining, much lower-probability, splits.
	}%
	\label{fig:root_splits}
\end{figure}

Analysis of the SARS-CoV-2 dataset of \citet{pekar2021timing} reveals notably better resolution at the root using HKY+RE than GTR.
In particular, we can examine the posterior support for different root splits (resolutions of the tree into two clades, or partitions of the taxa defined by the root).
The analysis with HKY+RE samples fewer potential resolutions of the root, and gives higher posterior probability to the most probable resolution (Figure~\ref{fig:root_splits}). 

\end{document}